%% file: main.tex
\documentclass[%
 aip,jap,
 amsmath,amssymb,
 reprint,%
]{revtex4-2}

\usepackage[acronym]{glossaries} 

\usepackage{graphicx}
\usepackage{dcolumn}
\usepackage{bm}
\usepackage[T1]{fontenc}
\usepackage{mathptmx}
\usepackage{braket}
\usepackage{booktabs}
\usepackage{tabularx} 

\usepackage{hyperref}
\hypersetup{colorlinks=true,linkcolor=blue,citecolor=blue,filecolor=blue,urlcolor=blue}

\bibpunct{[}{]}{,}{n}{}{,}
\providecommand{\keywords}[1]
{
  \small	
  \textbf{Keywords: } #1
}

\input{glossaries} 
\glsdisablehyper 

\begin{document}

\title{Quantum vs. classical: A comprehensive benchmark study for predicting time series with variational quantum machine learning} 

\author{Tobias Fellner}
\thanks{Corresponding author}
\email{tfellner@icp.uni-stuttgart.de}
\affiliation{Institute for Computational Physics, University of Stuttgart, 70569 Stuttgart, Germany}

\author{David A. Kreplin}
\email{david.kreplin@hs-heilbronn.de}
\affiliation{Fraunhofer Institute for Manufacturing Engineering and Automation, 70569 Stuttgart, Germany}
\affiliation{Heilbronn University of Applied Sciences, 74076 Heilbronn, Germany}

\author{Samuel Tovey}
\affiliation{Institute for Computational Physics, University of Stuttgart, 70569 Stuttgart, Germany}

\author{Christian Holm}
\affiliation{Institute for Computational Physics, University of Stuttgart, 70569 Stuttgart, Germany}


\begin{abstract}
Variational quantum machine learning algorithms have been proposed as promising tools for time series prediction, with the potential to handle complex sequential data more effectively than classical approaches. However, their practical advantage over established classical methods remains uncertain. In this work, we present a comprehensive benchmark study comparing a range of variational quantum algorithms and classical machine learning models for time series forecasting. We evaluate their predictive performance on three chaotic systems across 27 time series prediction tasks of varying complexity, and ensure a fair comparison through extensive hyperparameter optimization. Our results indicate that, in many cases, quantum models struggle to match the accuracy of simple classical counterparts of comparable complexity. Furthermore, we analyze the predictive performance relative to the model complexity and discuss the practical limitations of variational quantum algorithms for time series forecasting.

\end{abstract}


\maketitle

\input{content/01_introduction}
\input{content/02_01_models}
\input{content/02_02_data}
\input{content/02_03_training}
\input{content/03_results}
\input{content/04_discussion}
\input{content/05_conclusion}

\section*{Acknowledgments}
T.F., S.T., and C.H. gratefully acknowledge financial support from the German Research Foundation (Deutsche Forschungsgemeinschaft, DFG) under Germany’s Excellence Strategy EXC 2075-390740016. Furthermore, the authors acknowledge funding from the German Research Foundation (Deutsche Forschungsgemeinschaft, DFG) through the Compute Cluster grant no. 492175459. The authors thank the International Max Planck Research School for Intelligent Systems (IMPRS-IS) for supporting T.F. and C.H..
D.A.K acknowledges funding by the German Federal Ministry
of Economic Affairs and Climate Action through the project
AutoQML (grant no. 01MQ22002A). 
Parts of this work are based on the master thesis \textit{Quantum machine learning for time series prediction}~\cite{fellner_quantum_2024}.

The authors disclose the use of LLM-based tools for grammar and spelling.

\section*{Data availability statement}
The code to reproduce the results of this work is publicly available at \url{https://github.com/tobias-fllnr/VariationalQMLTimeSeriesBenchmark} and \url{https://doi.org/10.18419/DARUS-5559}.

\bibliography{main.bib}

\begin{appendix}
\input{content/appendix_01_models}
\input{content/appendix_02_qrnn_discussion}
\input{content/appendix_03_computaional_resources}
\input{content/appendix_04_datasets}
\input{content/appendix_05_convergence}
\input{content/appendix_06_scaling_seq_length}
\input{content/appendix_07_number_parameters}

\end{appendix}

\end{document}

%% file: glossaries.tex
\newacronym{qml}{QML}{quantum machine learning}
\newacronym{vqa}{VQA}{variational quantum algorithm}
\newacronym{qrnn}{QRNN}{quantum recurrent neural network}
\newacronym{rnn}{RNN}{recurrent neural network}
\newacronym{qlstm}{QLSTM}{quantum long-short term memory}
\newacronym{lstm}{LSTM}{long short-term memory}
\newacronym{qnn}{QNN}{quantum neural network}
\newacronym{pqc}{PQC}{parameterized quantum circuit}

%% file: content/01_introduction.tex
\section{Introduction}

The prediction of time series data is fundamental to making important decisions in fields such as finance, healthcare, and climate science. However, forecasting complex temporal patterns remains a challenge for classical models.
In recent years, \gls{qml}~\cite{wiebe_quantum_2016, schuld_introduction_2015, biamonte_quantum_2017} has emerged as a promising field that leverages quantum computing for machine learning tasks~\cite{havlicek_supervised_2019, huang_quantum_2022}. Despite its potential, the practical advantage of QML over classical methods remains uncertain. So far, demonstrations of quantum advantage in machine learning have been restricted to artificial problem settings~\cite{liu_rigorous_2021, huang_power_2021}.

Currently available quantum computers are constrained by limited qubit counts and hardware noise, which restricts the depth of quantum circuits. In this Noisy Intermediate-Scale Quantum (NISQ) era~\cite{preskill_quantum_2018}, \glspl{vqa} have attracted significant attention~\cite{cerezo_variational_2021}. In \gls{qml}, these algorithms are often implemented using \glspl{pqc}, which utilize parameter-dependent quantum gates to encode data and manipulate quantum states through trainable weights optimized via iterative updates to minimize a classical loss function~\cite{mitarai_quantum_2018}. 
Typically, encoding and parameterization are performed using single- and two-qubit Pauli rotation gates. The output of the model can be obtained by evaluating the expectation values of the quantum circuit. 
\Glspl{pqc} are particularly suited for NISQ devices and provide a versatile framework for various machine learning tasks~\cite{cerezo_challenges_2022}. Their application in classification has been extensively studied~\cite{havlicek_supervised_2019, mitarai_quantum_2018, schuld_quantum_2019, schuld_circuitcentric_2020}, and a wide range of quantum models for classification have been proposed~\cite{perez-salinas_data_2020, mari_transfer_2020, zoufal_variational_2021, lloyd_quantum_2020, zhang_trainability_2020}.

Recently, the ability of variational \gls{qml} models to process and learn sequential data has been demonstrated~\cite{takaki_learning_2021, li_quantum_2023, chen_quantum_2022, cao_linearlayerenhanced_2023, rivera-ruiz_time_2022}.
These models are primarily inspired by classical machine learning architectures for sequential data processing. 
For example, References~\cite{takaki_learning_2021} and~\cite{li_quantum_2023} propose \gls{qrnn} architectures that aim to adapt the idea of \glspl{rnn}~\cite{rumelhart_learning_1986, jordan_chapter_1997} to the quantum domain. 
Similarly, a \gls{qlstm} model has been proposed~\cite{chen_quantum_2022,cao_linearlayerenhanced_2023} representing the quantum analog of the classical~\gls{lstm} model~\cite{hochreiter_long_1997}.
Additionally, the potential of \glspl{qnn} for learning sequential data has been explored~\cite{rivera-ruiz_time_2022}.

While the studies above demonstrate that quantum models can make accurate predictions, it remains unclear how their performance compares to classical approaches and whether variational quantum models offer a distinct advantage for time series prediction.
Moreover, existing research predominantly targets one-step-ahead predictions within a given time sequence, a task often characterized by high linearity, as the predicted value often closely aligns with the sequence, making a linear continuation a reasonable approximation.
As a result, the ability of \gls{qml} models to handle more complex forecasting challenges, such as long-term predictions, remains uncertain. Addressing these gaps requires a comprehensive and minimally biased benchmark that includes challenging prediction tasks, allowing for a rigorous evaluation of quantum and classical models.

Given the wide range of classical machine learning methods developed for time series forecasting, benchmarking plays a crucial role in evaluating their effectiveness. Large-scale benchmarks have compared classical machine learning and linear models, highlighting the strengths of deep neural networks in time series prediction~\cite{ahmed_empirical_2010, masini_machine_2023}. Deep learning models, have also been extensively benchmarked for multi-step prediction tasks, demonstrating the predictive capabilities of different LSTM models~\cite{chandra_evaluation_2021}. Beyond predictive performance, studies have investigated evaluation strategies for time series forecasting~\cite{cerqueira_evaluating_2020} and benchmarked saliency-based interpretability methods for time series data~\cite{ismail_benchmarking_2020}.

Benchmarking classical models has been conducted across diverse domains, including medical~\cite{harutyunyan_multitask_2019, xie_benchmarking_2020}, financial~\cite{krollner_financial_2010, barra_deep_2020}, and energy consumption data~\cite{amalou_multivariate_2022}, as well as speech recognition~\cite{shewalkar_performance_2019}. In particular, chaotic dynamical systems have become widely used benchmarks for evaluating classical forecasting methods~\cite{fan_longterm_2020, ramadevi_chaotic_2022, shahi_prediction_2022a}. Their inherently complex, nonlinear dynamics make them particularly well-suited for testing the capabilities of time series machine learning models~\cite{gilpin_chaos_2021}.

In contrast, benchmarking efforts for \gls{qml} models are still in the early stages. A comprehensive benchmark for binary classification tasks has shown that classical models generally outperform quantum classifiers~\cite{bowles_better_2024}. In addition, a benchmark for quantum kernel methods has provided valuable insights into how the design of quantum models affects their performance~\cite{schnabel_quantum_2025}. Quantum reinforcement learning techniques have also been evaluated~\cite{kruse_benchmarking_2025, meyer_benchmarking_2025}. These quantum benchmarks challenge claims of quantum superiority and underscore the need for rigorous validation and clearer insights into how quantum principles contribute to quantum machine learning.
Despite the growing interest in QML models for time series prediction~\cite{takaki_learning_2021, li_quantum_2023, chen_quantum_2022, cao_linearlayerenhanced_2023, rivera-ruiz_time_2022}, no comparative analysis among the various models has been explored. Additionally, comparisons with different classical models for more complex prediction scenarios are currently absent in the literature.

\begin{figure}[t]
    \centering
    \includegraphics[width=\linewidth]{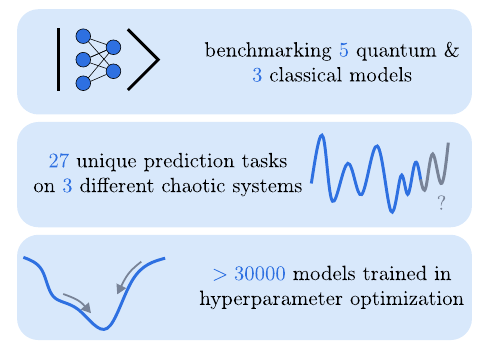}
    \caption{The key features of the benchmark study.}
    \label{fig:key_facts}
\end{figure}

In this work, we present, to the best of our knowledge, the first large-scale benchmark comparing variational quantum models with their classical counterparts for time series prediction. Our goal is to assess whether variational \gls{qml} can enhance performance on this task.
We evaluate five quantum and three classical machine learning models across 27 time series prediction tasks of varying complexity. Each model undergoes extensive hyperparameter optimization to ensure a fair comparison. A summary of the key features of this benchmark is provided in Figure~\ref{fig:key_facts}.
To establish an upper bound on their capabilities, quantum models are classically simulated under ideal, noiseless conditions using the Python library \texttt{PennyLane}~\cite{bergholm_pennylane_2022}.

The remainder of this paper is organized as follows. Chapter~\ref{chap:models} provides an overview of the quantum and classical machine learning models employed in this study. Chapter~\ref{chap:data} outlines the rationale behind the selection of prediction tasks, describes the datasets, and details the preprocessing steps. Training procedures and hyperparameter optimization strategies are presented in Chapter~\ref{chap:training}. The results are analyzed in Chapter~\ref{chap:results}, and their implications discussed in Chapter~\ref{chap:dis}. Finally, Chapter~\ref{chap:conclusion} summarizes the study and highlights potential directions for future research in \gls{qml} for time series analysis.

%% file: content/02_01_models.tex
\section{Model selection and implementation}
\label{chap:models}
\begin{figure*}
    \centering
    \includegraphics[width=1.0\linewidth]{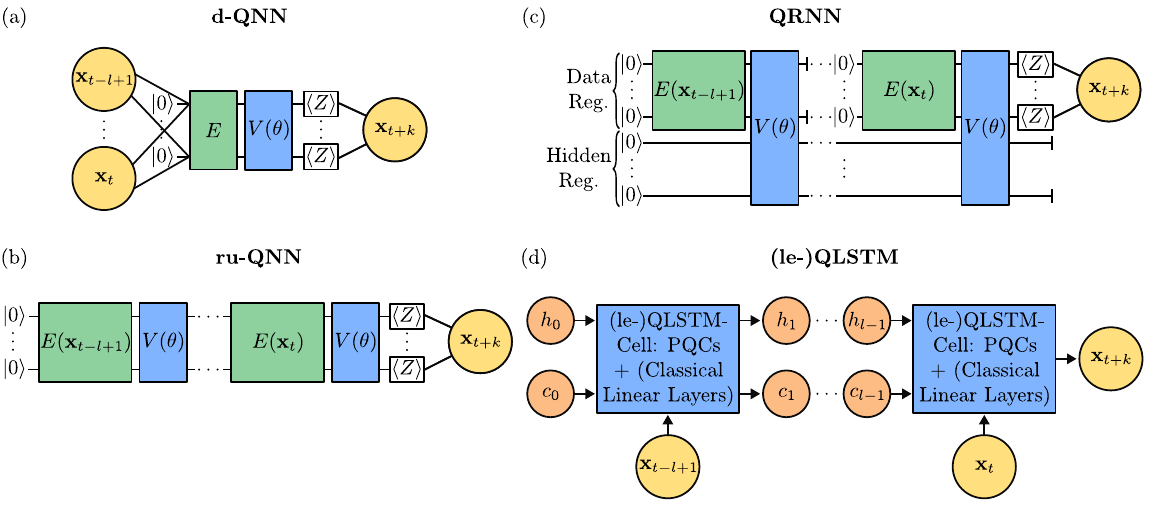}
    \caption{Overview of the different \gls{qml} models part of this study. $E$ and $V(\theta)$ denote the encoding layer and the variational layer, respectively. (a): In the d-QNN a \gls{pqc} is squeezed between two linear classical layers. (b): The ru-QNN utilizes the data re-uploading scheme by sequentially encodes data points. (c): The quantum circuit of the QRNN is divided into two registers: a data register, where sequential data encoding and measurement are performed, and a hidden register, which stores the hidden quantum state and passes it along through the network. (d): In the QLSTM model, the hidden states $h_i$ and cell states $c_i$ are passed between different cells. Each cell consists of \glspl{pqc} connected by a specific LSTM structure. In the case of the le-QLSTM, additional classical linear layers are incorporated within the cell.
}
    \label{fig:models_overview}
\end{figure*}

In this work, we consider a wide range of variational \gls{qml} models for time series prediction that have been previously proposed in the literature. We follow a similar procedure to~\cite{bowles_better_2024} and choose the quantum models by manually selecting relevant papers that introduce new variational quantum algorithms for time series prediction. By relevant papers, we mean papers with at least 25 citations as of June 30, 2025. While this selection criterion biases older publications, it ensures that the models to be benchmarked in the following have been influential in the \gls{qml} community. 

We found four papers that introduce different variational quantum models for time series forecasting. We group the models into three categories: Quantum neural networks, quantum recurrent neural networks, and quantum long short-term memory. In addition to the four models introduced by the selected papers, we benchmark an additional baseline QNN model that uses the data re-uploading scheme~\cite{schuld_effect_2021} combined with exponential encoding~\cite{shin_exponential_2023}. 
The quantum models that are part of this benchmark are briefly introduced in the following and shown in Figure~\ref{fig:models_overview}. Details on the circuit architecture, data encoding and readout, and hyperparameter values can be found in Appendix~\ref{app:models}. Unless otherwise noted, we use the identical architecture as proposed in the original papers. This choice aims to ensure a fair comparison avoiding arbitrary architectural biases.

\textbf{Dressed quantum neural network (d-QNN)}~\cite{rivera-ruiz_time_2022}: Based on the idea of a dressed quantum neural network~\cite{mari_transfer_2020}, the quantum circuit is squeezed between two trainable classical linear layers. The input layer transforms a time sequence to a dimension equal to the number of qubits. The resulting values are then encoded on individual qubits using angle encoding. After the trainable circuit ansatz and local single qubit Pauli-$Z$ expectation value measurements, the results are transformed by a classical linear output layer to the desired dimension of the target values. 

\textbf{Re-uploading quantum neural network (ru-QNN)}: This model follows the idea of the data re-uploading scheme~\cite{schuld_effect_2021}. A sequence of data points is encoded sequentially in the quantum circuit using exponential angle encoding~\cite{shin_exponential_2023}. This encoding strategy combined with the re-uploading structure results in a rich Fourier spectrum and therefore high expressivity. Variational blocks are inserted between these encoding blocks. At the end of the circuit, Pauli-$Z$ expectation values are measured and the resulting values are mapped to the desired target dimension by a classical linear layer. This model has not been introduced for time series prediction in the literature, but we include it as a baseline for comparison with other VQA models for time series prediction that do not make use of data re-uploading or exponential encoding.
Furthermore, we optimize the circuit ansatz of this model to derive a data-specific encoding circuit, and create a variational \gls{qml} tailored to the specific problem instance. In doing so, we establish an empirical upper bound for time series prediction using general QNNs. The procedure, which is based on a random search, is discussed in Chapter~\ref{chap:training} and Appendix~\ref{app:models}.

\textbf{Quantum recurrent neural network (QRNN)}~\cite{li_quantum_2023}: 
In this model, the quantum circuit is divided into a data register and a hidden register. The purpose of the data register is to encode sequential data and to obtain predictions. To achieve a recurrent structure,  the data register is reset to the initial zero state after each time-step. A variational layer then interleaves the data with the hidden register that carries the information of the time series in a quantum state along the sequence. 
The parameters of each variational layer are shared to achieve the recurrent structure. Finally, local single qubit Pauli-$Z$ expectation values of the data register are measured and mapped to the target dimension by a classical layer.
Simulating the reset of the qubits for variationally optimized circuits is challenging and restricted to tiny problem instances. 
Therefore, in Appendix~\ref{app:qrnn_reset_study}, we examine whether the reset of the data register is necessary at all.
For small systems, for which training with resets can be simulated, we find that resets to the initial state are not essential, and omitting the reset even improves performance. Consequently, the QRNN model used in our benchmark study has a modified structure compared to~\cite{li_quantum_2023}.

\textbf{Quantum long short-term memory (QLSTM)}~\cite{chen_quantum_2022}: 
A model based on the classical long short-term memory model~\cite{hochreiter_long_1997}. The LSTM structure consists of individual cells stacked on top of each other, each of which processes one data point of a sequence. Each cell consists of multiple neural network layers that are connected by a sophisticated structure to account for long- and short-term memory. A hidden state and a cell state are propagated from one cell to the next. In the quantum model, neural networks are replaced by \glspl{pqc}. The \glspl{pqc} are connected by a specific, fixed mathematical structure to ensure the transmission of a classical long- and short-term memory state between consecutive cells. The dimension of these hidden states depends on the number of qubits in the \glspl{pqc}.

\textbf{Linear layer enhanced QLSTM (le-QLSTM)}~\cite{cao_linearlayerenhanced_2023}: In addition to the QLSTM model, the \glspl{pqc} are connected by classical linear layers, by which the dimensions of the hidden state become independent of the number of qubits of the \glspl{pqc}.

\begin{table*}[ht]
\centering
\caption{Overview of the data sets and their relevant characteristics. A variety of data dimensions, mean periods, and Lyapunov times are provided to cover a wide range of different time series prediction problems. The prediction steps indicate how many steps ahead the models are trained to predict into the future. They are chosen to be approximately half or a full Lyapunov time as well as a single step ahead. The time unit is in terms of discrete time steps $t_{step}$ of the data sets. The plots display the evolution of the individual dimensions over time. For clarity, we restrict the visualizations to the first 15 mean periods of each data set.}
\begin{tabular}{lccc}
\toprule
 & \textbf{Mackey-Glass} & \textbf{Hénon map} & \textbf{Lorenz system} \\
\midrule
Dim. & 1 & 2 & 3 \\
Mean period & 44 & 4 & 19 \\
Lyap. time & 140 & 3.4 & 25 \\
Pred. steps & \{1, 70, 140\} & \{1, 2, 4\} & \{1, 13, 25\} \\
\midrule
Visualization &  & & \\
&
\includegraphics[width=0.2\textwidth]{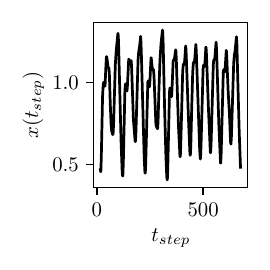} &
\includegraphics[width=0.2\textwidth]{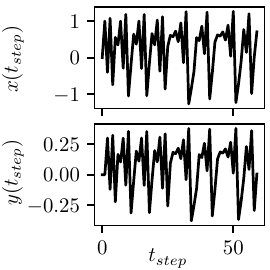} &
\includegraphics[width=0.2\textwidth]{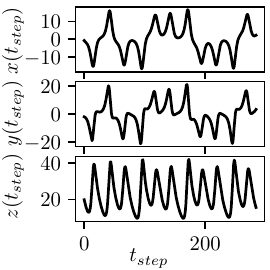} \\
\bottomrule
\end{tabular}
    \label{tab:datasets}
\end{table*}

For each of the three categories introduced above, we include a classical counterpart to the quantum models in order to set a baseline for the time series prediction capability of classical machine learning methods. As a classical analog for the QNN models, we choose a multi-layer perceptron (MLP) because it is consistent with the principle of a simple input-output architecture. For the QRNN and QLSTM models, we choose the classical counterparts, recurrent neural networks (RNN)~\cite{rumelhart_learning_1986, jordan_chapter_1997} and LSTM~\cite{hochreiter_long_1997}, on which these models are based. More details on the implementation can be found in Appendix~\ref{app:models}.
All quantum models are implemented and simulated using the \texttt{PennyLane} library~\cite{bergholm_pennylane_2022}. The classical models are implemented using the \texttt{PyTorch} library~\cite{paszke_pytorch_2019}. Preprocessing, training and postprocessing are done using \texttt{PyTorch}. Details on the computing resources used in this study can be found in Appendix~\ref{app:com_res}. 

When simulating \gls{qml} models, the number of qubits poses a significant limitation, as large systems are difficult to simulate classically, especially considering the extensive number of evaluations required for training.
Despite this challenge, classical simulations remain crucial, as executing and training these models on current quantum hardware is still limited by factors such as runtime, errors, and costs~\cite{kreplin_reduction_2024, buonaiuto_effects_2024, gujju_quantum_2024}.
The simulation of \gls{qml} models under error-free conditions serves as an upper bound of the predictive performance, providing insights into the relative performance of different quantum models and their competitiveness with classical methods.

%% file: content/02_02_data.tex
\section{Data selection and prediction task}
\label{chap:data}

A critical component of conducting a representative and meaningful benchmark study is the selection of appropriate learning problems. These problems must be sufficiently complex to allow models to learn advanced features beyond simple linear mappings or periodic correlations.
Consequently, this benchmark includes problems that extend beyond one-step-ahead predictions and employs data sets derived from dynamical systems exhibiting chaotic behavior. 
In the following, we will first discuss the considered task of time series prediction, before introducing the datasets used.

In this work, we deal with discrete time series data sets $\{\mathbf{x}_i\}_{i=1}^n, \mathbf{x}_i \in \mathbb{R}^d$.
We divide the data set into sequences of length $l$ consisting of consecutive data points, an approach known as the sliding window method~\cite{koc_analysis_1995}. The task of the models is to learn the mapping $f$ of these sequences onto a data point $\mathbf{x}_{t+k}$ that is $k$ steps ahead in the sequence
\begin{equation}
    f([\mathbf{x}_{t-l+1}, \dots, \mathbf{x}_{t}]) = \mathbf{x}_{t+k}\,.
\end{equation}
The difficulty of the learning task can be adjusted by varying the prediction step $k$. For small values of $k$, the problem often resembles a linear one, as the predicted value is close to the sequence. In such cases, a linear extrapolation of the data frequently serves as a good approximation. Increasing $k$ introduces more complexity into the problem, thereby increasing the difficulty of prediction.
For training and evaluation of the machine learning models, we end up with tuples consisting of sequences as inputs and their corresponding labels
\begin{equation}
\label{eq:tuples_sequence_label}
\begin{split}
    \Hat{X} =     \{([\mathbf{x}_1, \dots, \mathbf{x}_{l}], \mathbf{x}_{l+k}),\dots,
    ([\mathbf{x}_{n-k-l+1}, \dots, \mathbf{x}_{n-k}], \mathbf{x}_n)  \}\,.
\end{split}
\end{equation}
The sequence length is set to values of $l \in \{4, 8, 16\}$ for all models and all data sets in this benchmark.

\begin{table*}[ht]
    \centering
\caption{Overview of the models that are part of this benchmark study along with the model's hyperparameters. A detailed description of the models and the hyperparameter values can be found in Appendix~\ref{app:models}.\\
$^{\ast}$: The QRNN model part of the benchmarks study is the model with modified architecture compared to~\cite{li_quantum_2023} as motivated and described in Appendix~\ref{app:qrnn_reset_study}.}
    \begin{tabular}{c c}
        \textbf{Model} & \textbf{Hyperparameters} \\
        \midrule
        \midrule
        \textbf{Quantum models} & \\
        \midrule
         d-QNN & number of qubits, number of layers \\
         ru-QNN & number of qubits, circuit ansatz\\
         QRNN$^{\ast}$ & number of data qubits, number of hidden qubits\\
         QLSTM & number of qubits, number of layers\\
         le-QLSTM & number of layers, hidden size\\
         \midrule
         \textbf{Classical models} & \\
         \midrule
         MLP & number of layers, hidden size\\
         RNN & number of layers, hidden size\\
         LSTM & number of layers, hidden size\\
         \bottomrule
    \end{tabular}

    \label{tab:models_hyperparameters}
\end{table*}

To test the models against a variety of prediction challenges, we include different chaotic systems that differ in system dimensionality, mean period, and the time scale on which chaotic behavior occurs. 
The choice of chaotic systems is deliberate and multi-faceted. These systems are widely recognized in the machine learning literature as standard stress tests for non-linear temporal models due to their complexity and non-stationarity~\cite{fan_longterm_2020, ramadevi_chaotic_2022, shahi_prediction_2022a}. Unlike simpler time series, successfully predicting chaotic dynamics requires models to learn complex, high-dimensional non-linear mappings~\cite{gilpin_chaos_2021}, making them ideal for evaluating the capabilities of both classical and quantum machine learning architectures.

The data sets used are based on the one-dimensional delayed differential Mackey-Glass equation~\cite{mackey_oscillation_1977}, the two-dimensional Hénon map~\cite{henon_twodimensional_1976}, and the three-dimensional Lorenz system~\cite{lorenz_deterministic_1963}. 
The data sets and their characteristics are summarized in Table~\ref{tab:datasets}.
The time scale on which chaotic behavior occurs can be expressed in terms of the maximum Lyapunov exponent~\cite{wolf_determining_1985}.
This value quantifies the rate at which trajectories with two adjacent initial conditions diverge. The characteristic time scale on which chaotic behavior occurs is the inverse of the maximum Lyapunov exponent, which we call the Lyapunov time.
We are interested in this value in order to choose the prediction steps $k$ in such a way that the complex chaotic dynamics of the data must be learned by the models. Therefore we choose the prediction steps as approximately half and full Lyapunov times. Moreover, we include $k=1$, i.e. one-step-ahead,  for comparison. Details of the calculation of the Lyapunov times and mean periods as well as the mathematical details on the data sets can be found in Appendix~\ref{app:datasets}.

All data sets are obtained from the \texttt{ReservoirPy} library~\cite{trouvain_reservoirpy_2020}. Each data set consists of 1000 data points. We found this value to be a good compromise between having enough data to achieve generalization in training, while still being able to simulate the training of \gls{qml} models within a realizable time scale.

%% file: content/02_03_training.tex
\section{Training and hyperparameter tuning}
\label{chap:training}

In the following, we describe the training process as well as the hyperparameter tuning in this benchmark study. Each training consists of a specific data configuration (data set, sequence length, and prediction steps) and a model hyperparameter configuration. 
All data sets are scaled to the interval $[0, 1]$ using min-max scaling. For multi-dimensional data sets, each dimension is scaled independently. After rescaling, we construct a data set $\Hat{X}$ as described in Equation~\eqref{eq:tuples_sequence_label}. The first 60\% of $\Hat{X}$ are used as the training set, the next 20\% as the validation set, and the last 20\% as the test set. We train all models using the Adam optimizer~\cite{kingma_adam_2017} with a learning rate of 0.001, a batch size of 64, and the Mean Square Error (MSE) loss. Training is considered complete using a convergence criterion that monitors the loss on the validation data set (details in Appendix~\ref{app:convrgence}). As a measure of prediction accuracy we choose the MSE between the predicted and the target values.

To account for statistical variance, each hyperparameter setting is trained ten times with different random initial weights. We evaluate these runs using the median MSE on the validation set. To measure variability across runs, we report the median absolute deviation (MAD). Hyperparameter optimization over the different model hyperparameters listed in Table~\ref{tab:models_hyperparameters} is then done using a grid search to find the configuration that achieves the best median MSE on the validation set. This process is repeated for all models and data configurations.

To ensure comparability across models, we selected hyperparameter sets with a similar amount of values, ensuring that all models undergo a similarly complex hyperparameter optimization process. The specific hyperparameter values used in the benchmark are detailed in Appendix~\ref{app:models}. 
For classical models, hyperparameters were chosen to keep the number of trainable parameters roughly on the same order of magnitude as in quantum models.
Most models are optimized using a grid search over a 3×3 hyperparameter space for all data sets, sequence lengths, and prediction horizons.
However, certain models require more restricted hyperparameter searches due to computational constraints. The QLSTM model is computationally expensive to simulate. Therefore, optimization is limited to configurations with only two different qubit counts and three different layer counts. 
In contrast, the le-QLSTM model has three different model hyperparameters. To maintain consistency across models, the number of qubits is set to six for the le-QLSTM and not optimized throughout the benchmark.

The ru-QNN model is designed as a base-line QNN for time series prediction in this benchmark study.
It is known that the optimal variational ansatz for a given learning problem depends on the task to be learned~\cite{cerezo_variational_2021}.
Therefore, we optimize its ansatz as part of the hyperparameter optimization to create a data-specific \gls{qml} model for comparison. 
As the number of constructable ansätze is too large to optimize over the full space of possible ansätze, we follow a structured approach for ansatz optimization.
We construct random variational circuits by combining different encoding and variational blocks.
As demonstrated in Reference~\cite{rapp_reinforcement_2025}, a random circuit search is often sufficient to generate effective \gls{qml} models for a given data set.
The ansatz optimization is explained in Appendix~\ref{app:models}.

%% file: content/03_results.tex
\section{Benchmark results}
\label{chap:results}

\begin{figure*}
    \centering
    \includegraphics[width=0.8\linewidth]{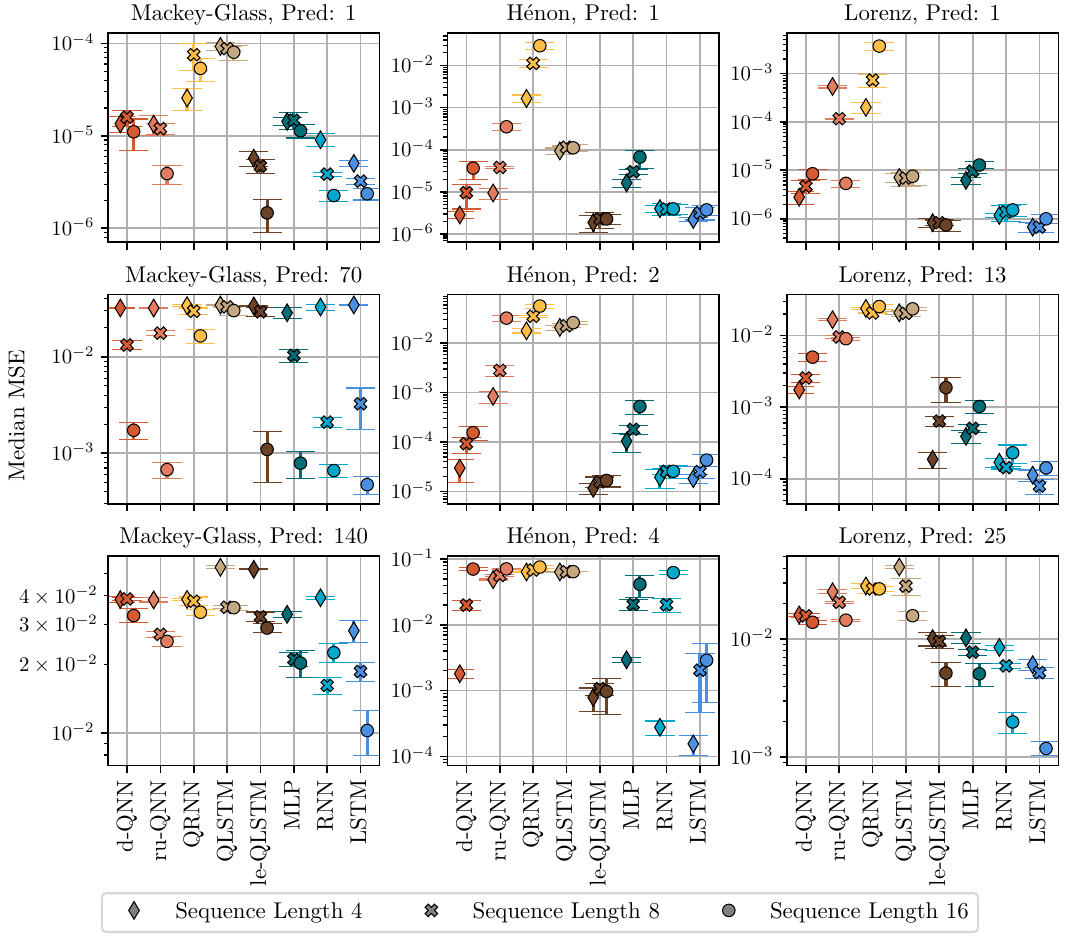}
    \caption{Benchmark results across combinations of data sets, prediction steps, and sequence lengths. The left column shows the results for the Mackey-Glass data set, the middle column shows the results for the Hénon data set, and the right column shows the results for the Lorenz data set. The top row shows the results for one-step ahead prediction, the middle shows the results for about half a Lyapunov time, and the bottom shows the results for about a full Lyapunov time. Within each subplot, the median MSEs for different sequence lengths are shown for each model, with different marker types indicating different sequence lengths. Error bars represent the MAD over ten random initializations. The results reflect the best performance of all model architectures and hyperparameters tested.}
    \label{fig:main_benchmark_plot}
\end{figure*}
In this section we discuss the outcomes of our benchmark study.
The results are shown in Figure~\ref{fig:main_benchmark_plot}. It displays the best median MSE for each optimized model on the test data set for each data set, number of prediction steps, and sequence length.
The prediction errors of the quantum models are shown on the left side of each subplot in red-brown colors, and the results of the three classical models are shown on the right side in blue colors. 

\subsection{Performance of the quantum models}
We first examine the two QNN models used in this study. 
The ru-QNN performs slightly better than the d-QNN on Mackey-Glass data, however, the d-QNN outperforms it on higher-dimensional data sets. On the Hénon data, the d-QNN exceeds the ru-QNN by several orders of magnitude over large prediction horizons. These findings are significant as we  developed the ru-QNN as a base-line QNN model for time series analysis that has been tailored to the data sets. 

A key difference is that the variational circuit of the d-QNN is embedded between classical layers. Therefore, it is possible that an important part of the training is done in the classical layers. There, the \gls{pqc} may serve only as an additional layer, contributing little to the time series prediction capability. This hypothesis is supported by the superior performance of the d-QNN on high-dimensional data. There, the number of classical optimizable parameters increase as the input and output are mapped by a linear layer that increases with data dimension. This scaling could explain the performance gap. However, heavy reliance on classical layers challenges the core goal of variational quantum algorithms, which is to exploit quantum properties such as exponential phase space scaling for potential advantages over classical methods. Such reliance may undermine this motivation.

Next, we examine the QRNN model. As motivated in Chapter \ref{chap:models} and discussed in more detail in Appendix ~\ref{app:qrnn_reset_study}, we here study an architecture that omits the reset of the qubits in the data register. Without this reset operation, the architecture effectively functions as a QNN, similar to the ru-QNN model.
In multi-step prediction tasks, both models exhibit similarly limited performance across data sets, likely due to their insufficient capacity to capture complex temporal patterns. However, in single-step prediction tasks, the ru-QNN models achieve significantly higher accuracy. This improvement is likely attributable to differences in the variational layer ansatz. Additionally, the QRNN architecture imposes a constraint on the number of qubits available for data encoding, as some qubits are reserved for carrying information across time steps. Allocating more qubits to the data encoding process may lead to a more expressive model, as suggested in \cite{schuld_effect_2021}.

Comparing the two QLSTM models, we find that the le-QLSTM consistently outperforms the original QLSTM of~\cite{chen_quantum_2022}, achieving at least an order of magnitude better accuracy across all data sets for prediction horizons up to half a Lyapunov time. This improvement likely stems from structural differences. The le-QLSTM introduces additional classical layers between individual \glspl{pqc} within the QLSTM cell, increasing both the number of trainable parameters and the size of the classical hidden states passed between cells. Unlike the ru-QNN and QRNN, both QLSTM models share the limitation that the hidden state is not propagated as a quantum state. Only measurement results of the quantum state are propagated after each \gls{pqc} within the QLSTM cell. This limitation may reduce their ability to exploit the exponentially large Hilbert space of quantum systems. Among all \gls{qml} models in the benchmark, the original QLSTM generally underperforms, while the le-QLSTM achieves the highest prediction accuracy in most cases, suggesting that its classical layers are primarily responsible for its success.

\subsection{Comparing to classical models}
The two quantum models with the best overall performance, especially for more complex tasks of predicting more than half a Lyapunov time, are the d-QNN and the le-QLSTM. Both models contain classical linear layers that enclose \glspl{pqc}, prompting the question of how much the quantum component contributes to learning, given the potential dominance of classical layers. To explore this, we compare the quantum models to established classical baselines.

Among classical models, the MLP performs the worst across nearly all tasks, likely due to its lack of considering the sequential structure of the data. The RNN and LSTM perform similarly on short horizons, with the LSTM slightly better for longer ones. These observations align with benchmarks performed on other data sets~\cite{amalou_multivariate_2022, shewalkar_performance_2019}. Taking the classical sequential models as baselines, we find that the best \gls{qml} models, namely the d-QNN and the le-QLSTM, achieve at most comparable prediction accuracies to the classical models. For prediction horizons of a full Lyapunov time, the classical models significantly outperform their quantum counterparts. These results question the usefulness for using VQAs for time series prediction. The QLSTM model shows inferior performance compared to its classical counterpart in the context of this study. Although the inclusion of linear layers in the architecture design, resulting in the le-QLSTM model, improves prediction accuracy, the overall performance of these models is below that of their classical counterpart. This observation suggests that there may be no discernible advantage to replacing classical neural layers within the LSTM architecture with quantum neural networks.
We also compare the d-QNN with the MLP. Both models share the way sequential data is injected into the models, as well as the structure of the input and output layers with an enclosed (quantum) layer. While the performance depends on the data set and the prediction horizon, overall the two models provide comparable prediction results. 
Our results raise the broader question whether adding variational quantum layers to classical models qualitatively improves their performance in the context of time series prediction. Alternatively, the quantum layers may simply act as substitutes for the classical layers, behaving similarly without offering a significant advantage.

Next, we investigate the scaling of prediction accuracy with respect to sequence length. In Appendix~\ref{app:scaling_sequ_length}, we show the benchmark results from Figure~\ref{fig:main_benchmark_plot} as a function of sequence length. We observe that the optimal input sequence length depends primarily on the data set and the prediction horizon. However, for a given learning problem, the optimal sequence length is largely identical for different models. While longer sequence lengths can improve prediction accuracy for the Mackey-Glass and Lorenz data sets, the Hénon data is best predicted with a sequence length of four. This may be related to the high mean frequency and short Lyapunov time of the Hénon data set. In this case, longer sequences may not facilitate the extraction of relevant features from the data and may instead obscure meaningful patterns by encoding additional information. Importantly, when comparing quantum and classical models, we observe no qualitative difference in their scaling behavior. This indicates that the comparative results of this benchmark are robust and likely hold when extrapolating to sequence lengths outside the range tested in this study.

\subsection{The role of parameter count on model performance}

\begin{figure}
    \centering
    \includegraphics[width=\linewidth]{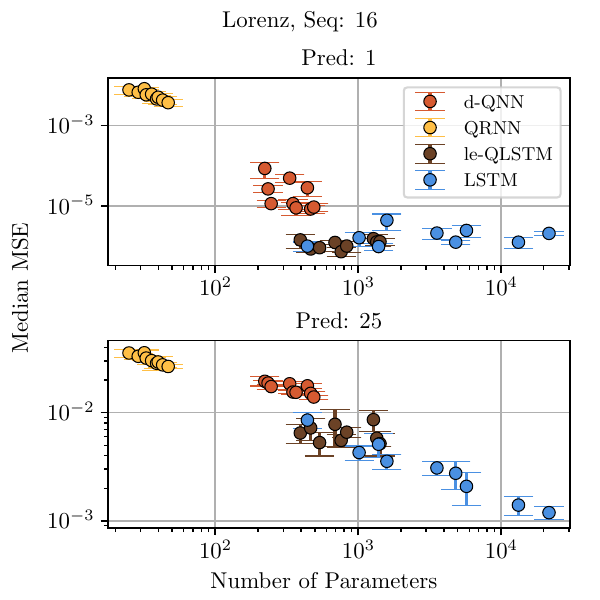}
    \caption{Here we show the median MSE over the number of parameters for different hyperparameter configurations of the d-QNN, le-QLSTM, QRNN, and LSTM models. The results are obtained for training on the Lorenz data set with a sequence length of 16. The upper plot shows the prediction errors for one-step prediction, while the lower plot shows the prediction errors for a prediction horizon of 25.}
    \label{fig:models_over_parameter}
\end{figure}
A critical consideration is that the different models in this study vary in the number of trainable parameters. By \textit{trainable parameters} we refer to all parameters that are optimized during training. For \gls{qml} models, this includes both classical parameters in linear layers and those embedded in quantum circuits. When selecting hyperparameter ranges, we ensure that the hidden states of the machine learning architectures are of comparable size and that their hyperparameters are optimized within ranges such that the number of trainable parameters are similar among all models to provide a fair comparison between models. Furthermore, we examine the relationship between the total number of trainable parameters and the predictive performance of the models.

Figure~\ref{fig:models_over_parameter} shows the median MSE from ten initializations on the test data set as a function of the number of trainable parameters for four machine learning models from the benchmark. The plot includes all hyperparameter configurations explored during the grid search, resulting in a number of models with different parameter counts. Error bars indicate the MAD of the test data set MSE. Results are based on the Lorenz data set with a sequence length of 16. The top plot shows one-step predictions, while the bottom plot shows predictions over a horizon of the more challenging 25 steps. These results are representative of other data sets, sequence lengths, and prediction horizons. To maintain clarity, we present only four models in the figure. Comprehensive results for all models are provided in Appendix~\ref{app:mse_over_parameters}.

For the subsequent analysis, we select the LSTM as a representative classical model due to its superior predictive performance in the benchmark. For quantum models, we choose the le-QLSTM and d-QNN, which showed the best forecasting results. Additionally, we include the QRNN, which inherently has a small number of parameters, to contextualize its performance relative to system size.

Figure~\ref{fig:models_over_parameter} shows that the number of trainable parameters of the QRNN is about an order of magnitude smaller than those of the other quantum models, which is due to the repeated structure and reused parameters for each data point of the time series. 
In contrast, the LSTM shows a comparable number of parameters to the d-QNN and le-QLSTM, but can also be up to an order of magnitude larger. Despite these size differences, for short prediction horizons (upper subplot), the LSTM already reached its best performance when the number of trainable parameters is comparable to the d-QNN and le-QLSTM. However, for longer prediction horizons (lower subplot), its performance continues to improve potentially because more parameters can be used during training as shown for the prediction horizon of 25. Nevertheless, when the LSTM is limited to sizes comparable to the d-QNN and le-QLSTM, it yields similar prediction accuracy.

An exception to the results in Figure~\ref{fig:models_over_parameter} is the performance of the le-QLSTM on the Hénon data set for a prediction horizon of four time steps. In this case, as shown in the corresponding figure in Appendix~\ref{app:mse_over_parameters}, the le-QLSTM achieves prediction accuracies comparable to those of the LSTM,  despite using up to an order of magnitude fewer parameters. This suggests that, for this specific task, the quantum layers in the le-QLSTM architecture may enhance time series prediction. However, since this effect is not observed for other data sets, this statement cannot be generalized to other learning tasks.

Overall, the results suggest that for more challenging prediction tasks, the best classical and quantum models show at most comparable performance when the number of trainable parameters is matched. The lower prediction accuracy of the QRNN is likely due to its limited number of trainable parameters. Although increasing the size of the QRNN by adding layers or qubits could improve its performance, the observed trends suggest that it is unlikely to outperform classical models of similar size.

%% file: content/04_discussion.tex
\section{Discussion}
\label{chap:dis}

\begin{figure}[t]
    \centering
    \includegraphics[width=\linewidth]{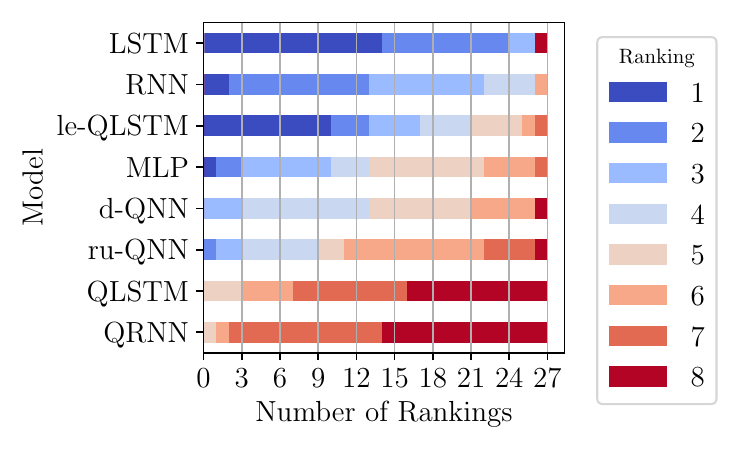}
    \caption{Ranking of the nine different models based on their best median MSE across all 27 learning problems. For each learning problem, the models are ranked according to their best median MSE after selecting the optimal architecture and hyperparameter set. Colors represent ranks, from dark blue (1st) to dark red (8th). Models are ordered according to the average rank obtained.}
    \label{fig:rankings}
\end{figure}

Similar to Reference~\cite{bowles_better_2024}, we rank the models for each prediction problem to provide a final overview of their performance across all data sets, prediction steps, and sequence lengths tested. This ranking allows for a direct comparison of their prediction capabilities. We show the results of this comparison in Figure~\ref{fig:rankings}. We see that the LSTM and the RNN achieve the best average rank, which is in line with our previous discussion. The third classical model, the MLP, is also among the models with an overall competitive performance. The best quantum models in this analysis are the le-QLSTM and the d-QNN, probably because learning can occur in the classical layers. Significantly, the two quantum models proposed specifically for processing temporal data, the QLSTM and the QRNN, perform the worst in this study. The ru-QNN, designed as a base-line QNN model, performs best among quantum models with few classical parameters. However, it still falls behind quantum models that contain a substantial number of classical parameters. Overall, our results call into question the benefit of using VQAs over classical machine learning methods for time series prediction.

While VQAs have a wide range of potential applications, they face a number of challenges and limitations~\cite{cerezo_challenges_2022} that apply to both classification and regression tasks. First, identifying the optimal ansatz of variational layer is a highly non-trivial problem that depends on the specific problem under consideration~\cite{bittel_optimal_2023, wu_efficient_2021}. Given the large number of potential configurations of gates that make up the ansatz, identifying an optimal one may require exponentially increasing resources. This could ultimately hinder the potential for a quantum advantage. In this work, we tackled this limitation by including an ansatz optimization of the ru-QNN model.

Another general limitation is the phenomenon of barren plateaus, where the loss landscape spanned by quantum parameters becomes exponentially flat as the number of qubits increases~\cite{mcclean_barren_2018, ragone_lie_2024}. This poses a challenge because exponentially more resources are required during training to find an optimal solution, and optimization becomes increasingly difficult.
The models studied in this work are specifically designed to be free from barren plateaus~\cite{cerezo_cost_2021}. However, recent research suggests that quantum models processing classical data while avoiding barren plateaus may become classically simulable~\cite{cerezo_does_2025}. This raises concerns that such models could be efficiently simulated with classical resources in polynomial time, potentially undermining a potential advantage of using \glspl{vqa} for classical data learning.
Nevertheless, the findings in this paper remain relevant to the quantum-inspired regime, where variational QML models operate efficiently on classical hardware.

The benchmark calculations in this work are performed under idealized conditions, without accounting for hardware limitations or finite-sampling noise. By simulating the models in this way, we aim to establish an upper bound on their performance. In real quantum hardware, several factors can introduce additional constraints~\cite{kreplin_reduction_2024, buonaiuto_effects_2024, gujju_quantum_2024}, including limited quality of qubits, environmental interactions, and finite sampling effects when estimating expectation values from probabilistic measurements. Consequently, when these hardware limitations come into play, the performance of the models discussed here is expected to degrade, leading to lower overall performance.

%% file: content/05_conclusion.tex
\section{Conclusion}
\label{chap:conclusion}

Designing a meaningful and informative benchmark study is a complex task that requires careful selection of prediction tasks, model architectures, and training methodologies. In this study, we tackled these challenges by systematically evaluating a diverse set of models on time series prediction tasks of varying complexity. To ensure a fair comparison, all models underwent comparable hyperparameter optimization and were trained using an identical procedure.

Our findings indicate that variational \gls{qml} methods generally perform at best on par with simple classical machine learning models across a broad range of prediction tasks. Notably, the quantum models that achieve the highest predictive performance often rely on a substantial number of classical trainable parameters, raising questions about the actual contribution of quantum effects in these models. Moreover, many VQAs specifically designed for time series processing, such as QLSTM and QRNN, tend to underperform compared to simpler QNN architectures in our benchmarks.

These results highlight the need for novel approaches that better exploit quantum resources for time series prediction as in variational methods. One promising direction is quantum reservoir computing~\cite{fujii_harnessing_2017, ghosh_quantum_2019, tovey_generating_2025}, which leverages the natural dynamics of quantum systems to process encoded sequences non-linearly. Unlike VQAs, this approach only requires training a classical output layer on quantum reservoir measurements, potentially overcoming limitations such as exponential concentration if designed appropriately~\cite{xiong_fundamental_2025}. Exploring such alternative strategies beyond \glspl{vqa} could open new avenues for advancing time series analysis using quantum resources.

%% file: content/appendix_01_models.tex
\section{Details on models}
\label{app:models}

\subsection*{Multi-layer perceptron}
In this study, we implement MLPs as one classical comparison for the quantum models. To perform time series prediction, a time series $[\mathbf{x}_{t-l+1}, \dots, \mathbf{x}_{t}]$ of length $l$ is fed into the input layer. In the case of a multi-dimensional time series with $d \geq 2$, the input vector is flattened. In general, the input layer maps a vector of size $l \cdot d$ to the size of the first hidden layer. Since there is considerable flexibility in the construction of the hidden layers, the number of hidden layers and the size of each hidden layer are hyperparameters of the model. In the benchmark, we evaluate a number of hidden layer configurations to identify regions in the hyperparameter space where the model performs well for a given problem. The hyperparameter values that are part of the grid search are shown in Table~\ref{tab:mlp_hyper}.
We use the rectified linear unit function as the activation. The output layer maps the last hidden state to the dimension $d$ of the data point in the time series to be predicted. Note that this architecture does not inherently account for the temporal structure of the data. However, it provides an initial baseline for the capabilities of classical neural networks against which quantum models can be compared in the benchmark study.

\begin{table}[htbp!]
    \centering
    \caption{Hyperparameters of the MLP models.}
    \begin{tabular}{cc}
        hyperparameter & values \\
        \hline
        number of layers & \{1, 2, 3\} \\
        hidden size & \{8, 16, 32\}
    \end{tabular}

    \label{tab:mlp_hyper}
\end{table}
\subsection*{Dressed quantum neural network~\texorpdfstring{\cite{rivera-ruiz_time_2022}}{}}

\begin{figure}[htbp!]
    \centering
    \includegraphics[width=\linewidth]{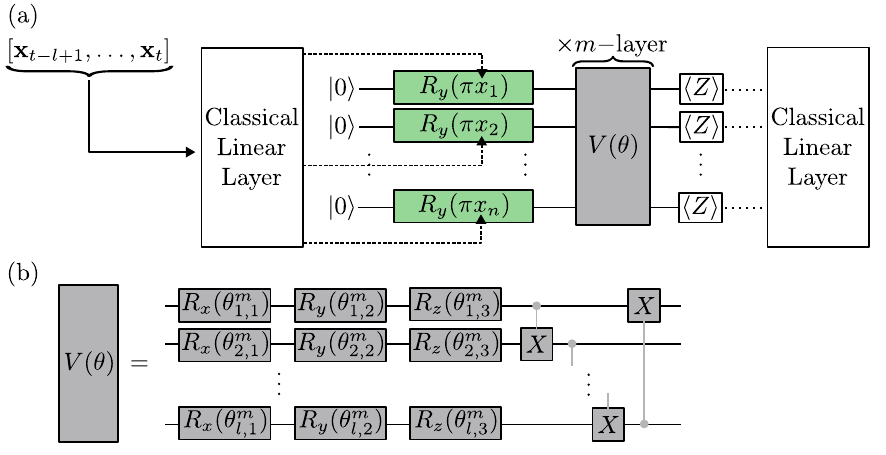}
    \caption{(a): Architecture of the d-QNN. The entire time sequence is first passed through the input layer and mapped to an $n$-dimensional representation, which is then encoded onto $n$ qubits. This approach decouples the number of qubits from the sequence length, making them independent parameters. (b): Circuit of a variational layer $V(\theta)$. Each layer has independent weights $\theta_{i,j}^m$.}
    \label{fig:d-vqc}
\end{figure}

The d-QNN model as introduced for time series prediction in~\cite{rivera-ruiz_time_2022} is based on the idea of a dressed QNN as proposed in~\cite{mari_transfer_2020}. In this approach a sequence $[\mathbf{x}_{t-l+1}, \dots, \mathbf{x}_{t}]$ is passed through a classical linear layer before encoding into the quantum system. The concept is illustrated in Figure~\ref{fig:d-vqc}. The sequence of length $l$ and dimension $d$ is mapped from the input dimension $l\cdot d$ to the target dimension $n$, which corresponds to the number of qubits. Each element in the target dimension is then encoded as a rotation around the $y$ axis in the Bloch sphere representation. Once the sequence is encoded in a quantum state, the variational layer $V(\theta)$ is applied. As suggested in~\cite{rivera-ruiz_time_2022}, the structure consists of $m$ such layers, where each layer contains rotation gates with weights $\theta_{i,j}^m$ followed by nearest neighbor entanglement $CNOT$ gates, as shown in Figure~\ref{fig:d-vqc}~(b).
First, the weights $\theta_{i,j}^m$ are randomly sampled from a uniform distribution in $[0,2\pi]$. After applying the variational layers $V(\theta)$, the quantum state is measured and the expectation values of the single qubit Pauli-$Z$ observables are obtained for all qubits. Unlike the approach in~\cite{rivera-ruiz_time_2022}, where only the Pauli-$Z$ expectation of the first qubit is measured to predict the next data point, we measure the expectation of all qubits separately. These results are then fed into a classical linear layer that maps the measurements to the dimensionality of the data point to be predicted. This ensures an equal and therefore more comparable treatment of data points with different dimensions.
Hyperparameters of the model are the number of qubits and the number of layers in the variational block. The values of the parameter scan are listed in the Table~\ref{tab:d-vqc_hyper}.

\begin{table}[htbp!]
    \centering
    \caption{Hyperparameters of the d-QNN models.}
    \begin{tabular}{cc}
        hyperparameter & values \\
        \hline
        number of qubits & \{4, 6, 8\} \\
        number of layers & \{1, 2, 3\}
    \end{tabular}
    \label{tab:d-vqc_hyper}
\end{table}
\subsection*{Re-uploading quantum neural network}

\begin{figure}[htbp!]
    \centering
    \includegraphics[width=\linewidth]{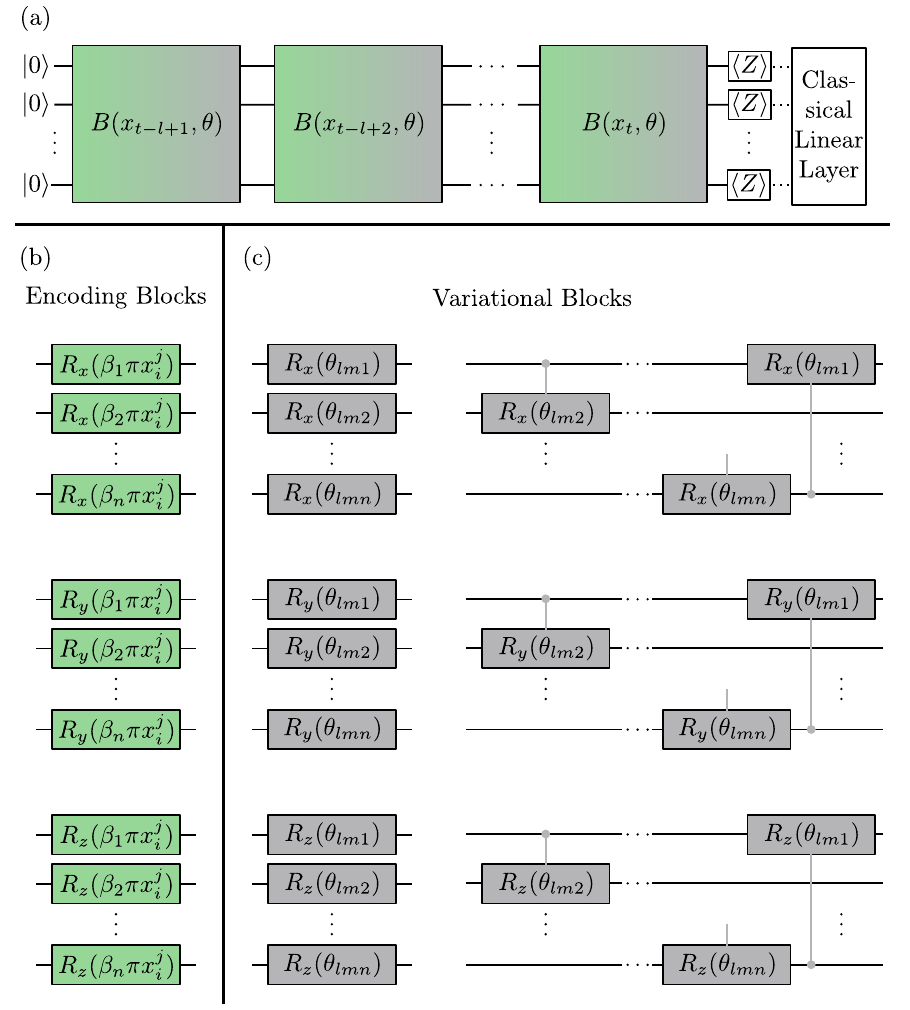}
    \caption{(a): The ru-QNN model. Data points of a sequence are encoded recurrently, with each point applied blockwise over all qubits. These encoding blocks are interspersed with variational layers $V(\theta)$.  The data reloading structure enhances the expressiveness of the circuit. (b): The encoding blocks encode the $i$-th data point of dimension $j$. The prefactors $\beta_a = 3^{a-1}/3^{n-1}$ are responsible for the exponential encoding, which allows access to a rich spectrum of Fourier frequencies. (c): Variational blocks can be separated into single-qubit blocks and two-qubits blocks with nearest neighbor entangling.}
    \label{fig:ru-vqc}
\end{figure}

We include an additional base-line QNN model for time series prediction. It is based on the data re-uploading scheme, which not only provides flexibility in choosing the sequence length and number of qubits independently, but also enhances the expressiveness of the quantum model~\cite{schuld_effect_2021}. 
The architecture is illustrated in Figure~\ref{fig:ru-vqc}~(a). Data points $\mathbf{x}_i$ in the sequence $[\mathbf{x}_{t-l+1}, \dots, \mathbf{x}_{t}]$ are encoded individually in each block $B$. We use exponential encoding~\cite{shin_exponential_2023}, which has been shown to extend the range of Fourier frequencies achievable, further increasing the expressiveness of the model. 

As we optimize the ansatz of the ru-QNN model, we randomly sample different ansätze as explained in the following. The ansätze are constructed by assembling different blocks of quantum gates. These blocks can be divided into encoding and variational blocks. The different blocks are shown in Figure~\ref{fig:ru-vqc}~(b)~and~(c). 
For encoding, we randomly draw one to three encoding blocks per dimension of the data to be learned. Note that the same block can be drawn multiple times. Similarly, we choose one to twelve variational blocks. In this way, we end up with randomly drawn sets of blocks, where the smallest possible set contains two blocks (one encoding block for one-dimensional data plus one variational block) and the largest possible set contains 21 blocks (three encoding blocks for each dimension of three-dimensional data plus 12 variational blocks).

The set of blocks is randomly ordered, but must obey the following constraints:
\begin{itemize}
    \item The first block must not be an entanglement block, as this would have no effect in the first time step, since the initial state is the zero state.
    \item The last block must not be a block consisting of Pauli-$Z$. Since we are measuring Pauli-$Z$ operators, such a block would not affect the measurements.
    \item Two consecutive blocks cannot have the same Pauli operator. This includes the last and the first block, since they are consecutive blocks when the solutions are added in the final model. The reason for this restriction is that consecutive blocks with rotations around the same axis can become redundant.
\end{itemize}
If no ansatz can be found that satisfies all the constraints for a particular set of sampled blocks, that set is omitted.

For each individual learning problem we randomly sample and train 100 variational circuits, each with a single random parameter initialization and three different numbers of qubits (4, 6, and 8).
Out of the resulting 300 models trained for each learning task, we determine the ten models with the best MSE on the validation set.
These ten ansätze are subsequently trained for ten random initial weights to obtain the final median MSE. The model with the best median MSE is the optimized ru-QNN for the given learning task.

The values of the hyperparameters are shown in Table~\ref{tab:ru-vqc_hyper}.

\begin{table}[htbp!]
    \centering
    \caption{Hyperparameters of the ru-QNN models.}
    \begin{tabular}{cc}
        hyperparameter & values \\
        \hline
        number of qubits & \{4, 6, 8\} \\
        circuit ansatz & \{100 random circuits\}
    \end{tabular}
    \label{tab:ru-vqc_hyper}
\end{table}
\subsection*{Recurrent neural network~\texorpdfstring{\cite{rumelhart_learning_1986, jordan_chapter_1997}}{}}
We use the \texttt{PyTorch} implementation of the RNN within this study. The activation function used is the hyperbolic tangent. Hidden size and number of layers are hyperparameters of the model. The values used for the grid search are shown in Table~\ref{tab:rnn_hyper}.

\begin{table}[htbp!]
    \centering
    \caption{Hyperparameters of the RNN models.}
    \begin{tabular}{cc}
        hyperparameter & values \\
        \hline
        number of layers & \{1, 2, 3\} \\
        hidden size & \{8, 16, 32\}
    \end{tabular}
    \label{tab:rnn_hyper}
\end{table}

\subsection*{Quantum recurrent neural network~\texorpdfstring{\cite{li_quantum_2023}}{}}

\begin{figure}[htbp!]
    \centering
    \includegraphics[width=\linewidth]{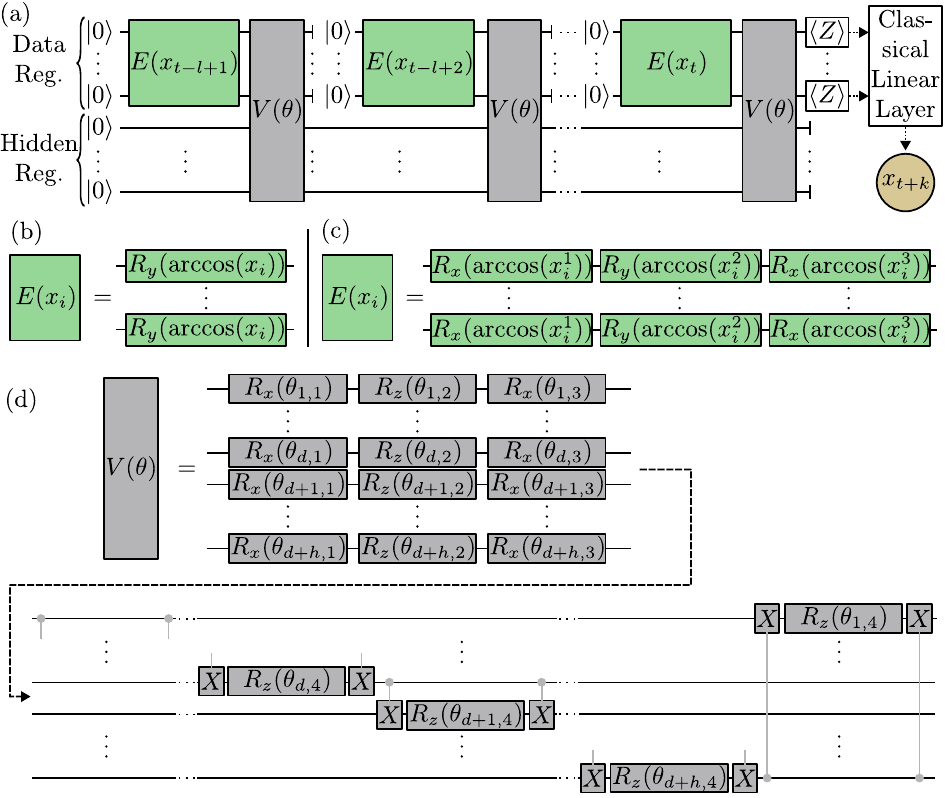}
    \caption{(a): Quantum recurrent neural network architecture. The circuit is divided into a data register and a hidden register. The sequence is recurrently encoded into the data register. After each encoding block, a variational layer is applied to facilitate the model's learning of the mapping from the quantum state in the data register to the hidden register. After the last point in the sequence is encoded, the data register is measured and the result is passed through a classical linear layer to predict the next data point. (b): Encoding circuit for one-dimensional data. (c): Encoding circuit for three-dimensional data. (d): Concept of the variational layer used throughout the thesis.}
    \label{fig:qrnn}
\end{figure}

The QRNN extends the concept of sequential learning in a QNN by assigning $d$ qubits to a data register and $h$ qubits to a hidden register. In this setup, a sequence of data points is encoded into the data register in a recurrent manner. The hidden register is inspired by the classical RNN, where a hidden state is transferred from one data point to the next. In the QRNN, the quantum state of the qubits in the hidden register serves as the state that is passed along the sequence. The QRNN architecture as introduced in~\cite{li_quantum_2023} is shown in Figure~\ref{fig:qrnn}~(a). Similar to a classical RNN, the QRNN uses a blockwise recurrent structure. In each block, a data point $\mathbf{x}_i$ from the sequence $[\mathbf{x}_{t-l+1}, \dots, \mathbf{x}_{t}]$ is encoded onto the data register. In the case of one-dimensional data, a data point $x_i$ is encoded by angle encoding using a Pauli-$Y$ rotation gate $R_y(\arccos(x_i))$ on all $d$ qubits of the data register. The corresponding circuit is shown in Figure~\ref{fig:qrnn}~(b). To ensure comparability with the original paper, we scale the input using the arccosine function as suggested in~\cite{li_quantum_2023}. Since the data is scaled to the interval $[0,1]$, the angle of rotation is within $[0,\pi/2]$. In the case of three-dimensional data, the three dimensions of a data point are encoded sequentially with rotation gates on each qubit by $R_x(\arccos(x_i^3))R_y(\arccos(x_i^2)R_x(\arccos(x_i^1))$.
The circuit for this encoding is shown in Figure~\ref{fig:qrnn}~(c). 
For two-dimensional data, the last layer of Pauli-$X$ rotation gates in each encoding block is omitted. After encoding the data point into the data register, a variational layer is applied. This layer contains trainable weights $\theta_{i,j}$ and entangles qubits from the data register with those in the hidden register. This entanglement allows the model to learn how to transfer the quantum representation of the data into the hidden state, facilitating the extraction of relevant features from the sequence. The specific approach used in~\cite{li_quantum_2023} and throughout this work is shown in Figure~\ref{fig:qrnn}~(d). After a layer of three parameterized rotation gates, a layer of nearest neighbor entanglement elements is applied. Each element consists of a parameterized Pauli-$Z$ rotation gate placed between two $CNOT$ gates. It is important to note that the weights of the variational layers are shared across all cells of each time step, similar to the RNN approach. 
After the block encoding the last data point of a sequence, the data register is measured. The single qubit Pauli-$Z$ expectation values of all $d$ qubits are passed into a classical linear layer to map to the dimension of the data point $\mathbf{x}_{t+k}$ to be predicted.
In the originally proposed model, after each of the first $l-1$ blocks, the quantum state of the data register is reset to the ground state $\ket{0}$ in preparation for initializing the next data point. Since resetting qubits is computationally expensive in the pipeline built for this work using \texttt{PennyLane}, we compare this originally proposed approach with one that omits resetting after each block. In this case, the quantum state of the data register propagates along the sequence instead. In chapter~\ref{chap:results} we show the results of this study for a small number of qubits and a small sequence length. For more qubits and longer sequence lengths, training is performed exclusively on architectures that do not incorporate the data qubit reset.
The hyperparameters of the architecture include the number of qubits in the data register and the number of qubits in the hidden register. Both approaches are trained with two qubits in each register. The hyperparameter ranges for the approach without reset are listed in Table~\ref{tab:qrnn_hyper}.

\begin{table}[htbp!]
    \centering
    \caption{Hyperparameters of the QRNN models without resetting the data register after each block. The architecture where the qubits of the data register are reset after each block was only trained for two qubits in each register.}
    \begin{tabular}{cc}
        hyperparameter & values \\
        \hline
        number of data qubits & \{2, 3, 4\} \\
        number of hidden qubits & \{2, 3, 4\}
    \end{tabular}
    \label{tab:qrnn_hyper}
\end{table}
\subsection*{Long short-term memory~\texorpdfstring{\cite{hochreiter_long_1997}}{}}
We use the \texttt{PyTorch} implementation of the LSTM. The hyperparameter values for the number of layers and the hidden sizes are given in Table~\ref{tab:lstm_hyper}.

\begin{table}[htbp!]
    \centering
    \caption{Hyperparameters of the LSTM models.}
    \begin{tabular}{cc}
        hyperparameter & values \\
        \hline
        number of layers & \{1, 2, 3\} \\
        hidden size & \{8, 16, 32\}
    \end{tabular}
    \label{tab:lstm_hyper}
\end{table}
\subsection*{Quantum long short-term memory~\texorpdfstring{\cite{chen_quantum_2022}}{}}

\begin{figure}[htbp!]
    \centering
    \includegraphics[width=\linewidth]{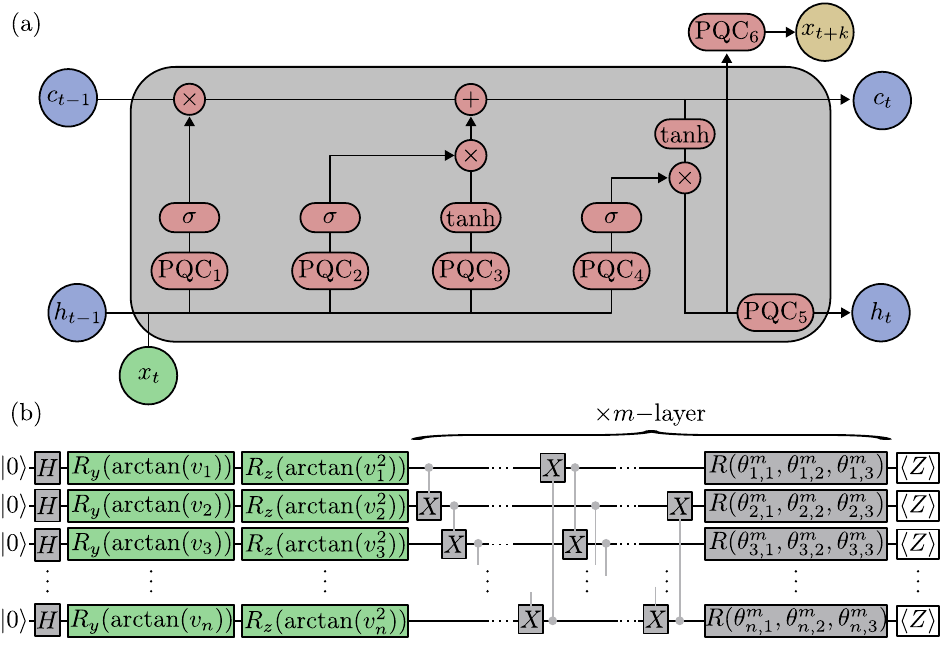}
    \caption{(a): Architecture of a QLSTM cell. In this design, the classical neural networks used in a traditional LSTM cell are replaced by \glspl{pqc}. Two additional \glspl{pqc} are incorporated: one to adjust the dimensionality of the hidden state and another to process the cell output to ensure that it is appropriately mapped to the data point to be predicted. (b): Schematic of the \glspl{pqc}. The input vector $\mathbf{v}_t$ is encoded using angle encoding, scaled by an arc tangent function. This is followed by the application of a layered variational block. Each layer consists of nearest and second-nearest neighbor entanglement operations using $CNOT$ gates, followed by a set of parameterized single-qubit rotation gates.}
    \label{fig:qlstm}
\end{figure}

We now discuss the quantum analog of a LSTM, known as Quantum Long Short-Term Memory (QLSTM), first proposed in~\cite{chen_quantum_2022}. The core idea of the QLSTM is to replace the classical neural network layers with \glspl{pqc}. The architecture of a QLSTM cell is shown in Figure~\ref{fig:qlstm}~(a). In this architecture, the four neural networks of the classical LSTM are replaced by PQC$_1$ to PQC$_4$. 
As in the classical LSTM, the previous hidden state $h_{t-1}$ and the current data point $\mathbf{x}_t$ of the sequence with dimension $d$ are concatenated into a vector $\mathbf{v}_t$. The dimension of this vector is equal to the number of qubits $n$ in all \glspl{pqc}. Thus, the dimension of the hidden state is given by $h=n-d$. The vector $\mathbf{v}_t$ is processed through PQC$_1$ to PQC$_4$, where the expectation value of the Pauli-$Z$ observable is measured on each qubit of each \gls{pqc}. The outputs of these \glspl{pqc} are then fed into the internal QLSTM structure, which mirrors the classical LSTM architecture. Since the outputs of PQC$_1$ to PQC$_4$ are multiplied element-wise by the cell state $c_t$, the dimension of $c_t$ is also equal to the number of qubits $n$.
To pass a hidden state $h_t$ to the next cell, the dimension must be reduced from $n$ to $h$. This is done by introducing an additional PQC$_5$. Although its approach is identical to the other \glspl{pqc}, only the expectation values of the first $h$ qubits are measured and propagated to the next QLSTM cell. As in the classical LSTM, these cells are stacked so that the hidden state $h_t$ and the cell state $c_t$ propagate through the sequence. When the last data point in the sequence is reached, an additional PQC$_6$ is applied. The internal $n$-dimensional state is passed through PQC$_6$, where the expectation values of all qubits are measured. Finally, a linear layer maps the measurement results to the dimension of the predicted data point $\mathbf{x}_{t+k}$.
The circuit architecture of all \glspl{pqc} is identical and is shown in Figure~\ref{fig:qlstm}~(b) as introduced in~\cite{chen_quantum_2022} and implemented in this work.  First, a Hadamard gate $H$ is applied to all qubits, followed by encoding the $i$-th element of the input vector $\mathbf{v}_t$ onto the $i$-th qubit via $R_z(\arctan(v_i^2)) R_y(\arctan(v_i))$.
Since the data is scaled to the interval $[0,1]$, the rotation angles for encoding are within $[0,\pi/4]$ for PQC$_1$. After the encoding layer, a series of $m$ variational layers are applied. Each layer consists of nearest and second-nearest $CNOT$ entanglements, followed by parameterized rotation gates defined as $R(\theta_{i, 1}^m, \theta_{i,2}^m, \theta_{i,3}^m) = R_z(\theta_{i,3}^m) R_x(\theta_{i,2}^m) R_z(\theta_{i,1}^m)$ for each qubit $i$. The weights $\theta_{i,j}^m$ are initialized by uniformly sampling from $[0,2\pi]$ before training. These parameters are independent across different PQC$_i$, but remain identical for the same PQC$_i$ across different QLSTM cells. Therefore, the total number of trainable quantum parameters is $6 \cdot 3 \cdot n \cdot m$. 
The hyperparameters that are part of the hyperparameter optimization in this work are the number of qubits $n$ as well as the number of layers $m$ in the \gls{pqc} ansatz and are listed in Table~\ref{tab:qlstm_hyper}. As simulating the QLSTM training with 8 qubits exceeds our computational resources, we only perform the hyperparameter optimization over 4 and 6 qubits.

\begin{table}[htbp!]
    \centering
    \caption{Hyperparameters of the QLSTM models.}
    \begin{tabular}{cc}
        hyperparameter & values \\
        \hline
        number of qubits & \{4, 6\} \\
        number of layers & \{1, 2, 3\}
    \end{tabular}
    \label{tab:qlstm_hyper}
\end{table}
\subsection*{Linear layer enhanced quantum long short-term memory~\texorpdfstring{\cite{cao_linearlayerenhanced_2023}}{}}

\begin{figure}[htbp!]
    \centering
    \includegraphics[width=\linewidth]{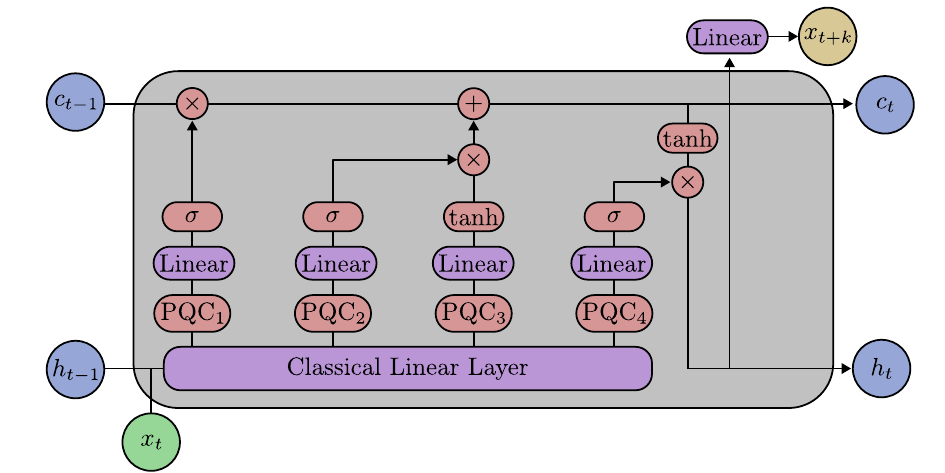}
    \caption{Structure of the le-QLSTM cell. To decouple the dimensions of the hidden and cell states from the number of qubits in the \glspl{pqc}, a series of classical linear layers are inserted. A linear input layer first maps the input vector $\mathbf{v}_i$ to a dimension corresponding to the number of qubits $n$. Following the \glspl{pqc}, another set of linear layers then maps the $n$ measurement results to the dimension of the hidden states.}
    \label{fig:le-qlstm}
\end{figure}

One limitation of the QLSTM architecture is that the size of the hidden state $h=n-d$ and the cell state $c=n$ both scale with the number of qubits $n$. Since the hidden and cell states are meant to store information about the sequence, it would be advantageous to have more flexibility in choosing these sizes. 
To decouple the hidden state size from the number of qubits, an advanced version of the QLSTM has been proposed in~\cite{cao_linearlayerenhanced_2023}. This variant, called linear layer enhanced quantum long short-term memory (le-QLSTM), incorporates classical linear layers into the architecture, as shown in Figure~\ref{fig:le-qlstm}. An input linear layer maps the input vector $\mathbf{v}_t$ to the number of qubits $n$ in PQC$_1$-PQC$_4$. After processing these \glspl{pqc}, additional linear layers map the number of qubits $n$ to the hidden and cell state size $h=c$. This eliminates the need for PQC$_5$ as in the original QLSTM architecture. Similarly, PQC$_6$ is replaced by a linear layer that maps the hidden state size $h$ to the dimension $d$ of the predicted data point. In~\cite{cao_linearlayerenhanced_2023}, as well as in the benchmarks performed, the circuit design remains identical to that shown in Figure~\ref{fig:qlstm}~(b). While this approach effectively decouples the number of qubits $n$ from the dimensions of the hidden and cell states $h$ and $c$, it also increases the number of classical parameters.
The hyperparameters of this model are the hidden sizes $h$ of the classical layers as well as the number of qubits $n$ and the number of layers $m$ in the \gls{pqc}. Since all other models in this benchmark have only two tunable hyperparameters to perform hyperparameter optimization, we set $n$ to the maximum of $n=6$ in this study. This ensures that all models in this benchmark undergo comparable hyperparameter optimization. The ranges of the other hyperparameters are shown in the Table~\ref{tab:le-qlstm_hyper}.

\begin{table}[htbp!]
    \centering
    \caption{Hyperparameters of the le-QLSTM models.}
    \begin{tabular}{cc}
        hyperparameter & values \\
        \hline
        number of layers & \{1, 2, 3\} \\
        hidden size & \{8, 16, 32\}
    \end{tabular}
    \label{tab:le-qlstm_hyper}
\end{table}

%% file: content/appendix_02_qrnn_discussion.tex
\section{Analysis of qubit reset in the QRNN model}
\label{app:qrnn_reset_study}
\begin{figure}[t]
    \centering
    \includegraphics[width=0.7\linewidth]{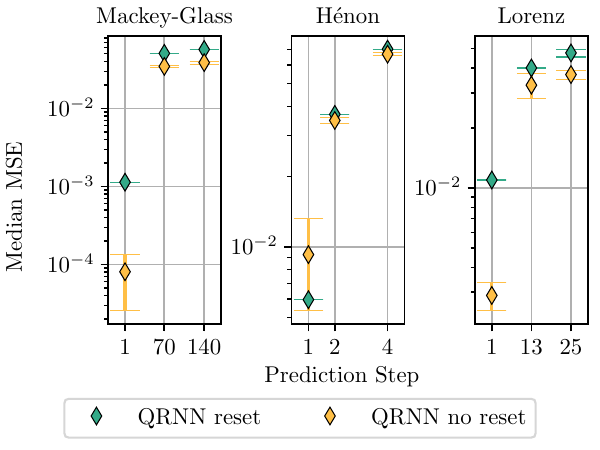}
    \caption{Comparison of the best median MSE obtained by the two QRNN architectures, with and without qubit resets in the data register. The results are for systems with two qubits in the data register and two qubits in the hidden register. All results shown here are obtained for a sequence length of four.}
    \label{fig:qrnn_rest_comparison}
\end{figure}

With the pipeline built for this study, efficient simulation of QRNN training as introduced in~\cite{li_quantum_2023} is not possible for all prediction problems. In \texttt{PennyLane}, as well as in most other quantum circuit simulators, efficient gradient calculation is not possible when resetting individual qubits to the $\ket{0}$ state, and instead additional qubits are swapped in during the simulation. 
Therefore, training the QRNN model where the data qubits are reset is in this framework only possible for a small number of qubits and a small sequence length. We note however that alternative approaches to circumvent this issue like density matrix emulation exist~\cite{viqueira_density_2025}.
To determine the effect of resetting the data qubits on model performance, we train identical models of QRNNs with and without resetting the quantum state of the data qubits to $\ket{0}$.
We carry out this study for two qubits in the data register and two qubits in the hidden register and a sequence length of four for all data sets and prediction lengths of this benchmark. 
Due to the additional qubits that are necessary for simulating the reset of the data qubits the total number of qubits for this simulation is ten qubits. 
The results of this study are shown in Figure~\ref{fig:qrnn_rest_comparison}. 
It becomes clear that for most of the prediction problems studied here, both architectures lead to similar results. Omitting the qubit reset leads to slightly better prediction accuracy in most cases. We interpret this finding as follows: By resetting the qubits in the data register after each sequence step, the information of previous steps contained in the data qubit subsystem is effectively lost. Only the hidden qubit register can carry information about past time steps, which can ultimately be used in training. While there is no guarantee that these observations will scale to models with more qubits, our results question the rationale for performing data qubit resets as introduced in~\cite{li_quantum_2023}. 
However, we have to note that by omitting the reset, the model effectively becomes a QNN, similar to the ru-QNN part of this study. 

%% file: content/appendix_03_computaional_resources.tex
\section{Details on computational resources}
\label{app:com_res}

All preprocessing, training and postprocessing was performed on Central Processing Units (CPUs). Training was done on a cluster with 18 nodes (2x32 cores, 3.85~GHz, 320~W and 384~GB RAM). Each individual training of a model was performed as a single core job. Most jobs a RAM of 2~GB is allocated, however for simulation systems with eight qubits 4~GB are used. The QRNN model with resetting the data qubits is computationally more expensive, why 8~GB are used. The node dwell time of the cluster used is 48 hours. Models are chosen in such a way that the models converge within that time. 

%% file: content/appendix_04_datasets.tex
\section{Details on the data}
\label{app:datasets}

\subsection{Determining the Lyapunov times}
Here we describe how we determine the Lyapunov exponents of the chaotic data sets used in this work. We use the algorithm of Rosentein et al.~\cite{rosenstein_practical_1993} to estimate the largest Lyapunov exponent of a discrete data set. We use the implementation of the \texttt{nolds} library~\cite{scholzel_nonlinear_2019}. The algorithm depends on two parameters \texttt{min\_tsep} and \texttt{lag}. As suggested in the original paper, we choose \texttt{min\_tsep} as the mean period of a signal and \texttt{lag} as the distance where the autocorrelation function falls below $1 - 1/e$ times its maximum value. For multi-dimensional data, we determine \texttt{min\_tsep} and \texttt{lag} separately for each dimension. We use the found values to compute the Lyapunov exponents using the algorithm~\cite{rosenstein_practical_1993}. Since the algorithm contains random elements, we average over 100 runs to determine the Lyapunov exponent. For multi-dimensional data, we take the average of the different dimensions to determine the Lyapunov exponent of the time series data set. The inverse is then the Lyapunov time, which is used as a measure of the timescale at which chaotic behavior occurs in the data used in this benchmark. 

\subsection{Determining the mean periods}
To obtain the mean period of a data set, the frequency spectrum is computed using the Fast Fourier Transform. The mean frequency is calculated as the amplitude-weighted average of the positive frequency components, and the mean period is then given by the inverse of this mean frequency.

\subsection{Chaotic data sets}
\subsubsection{Mackey-Glass equation~\texorpdfstring{\cite{mackey_oscillation_1977}}{}}
The Mackey-Glass equation is a one-dimensional delayed differential equation:
\begin{equation}
\label{eq:mackey_meth}
    \frac{\mathrm{d} x(t)}{\mathrm{d}t} = \frac{\alpha x(t-\Tilde{t})}{1 + x(t-\Tilde{t})^n} - \gamma x(t)\,.
\end{equation}
Here, $\alpha$, $\gamma$, $n$, and $\Tilde{t}$ are system parameters. The solution $x(t)$ describes the Mackey-Glass time series. We choose $\alpha = 0.2$, $\gamma = 0.1$, $n = 10$, and $\Tilde{t} = 17$, for which chaotic behavior occurs. We set the initial condition to $x(t=0) = 1.2$. The solution is obtained by \texttt{ReservoirPy} using a Runge-Kutta method~\cite{runge_ueber_1895, kutta_beitrag_1901} with a step size of 1.

\subsubsection{Hénon map~\texorpdfstring{\cite{henon_twodimensional_1976}}{}}
The Hénon map is a two-dimensional system described by the equations 
\begin{equation}
\begin{split}
    x_{n+1} &= 1 - ax_{n}^2 + y_{n}\,, \\
    y_{n+1} &= bx_{n}\,.
\end{split}
\end{equation}
We use the parameters $a=1.4$ and $b=0.3$ for which the map shows chaotic behavior. We set the initial condition to $x_0=y_0=0$.

\subsubsection{Lorenz system~\texorpdfstring{\cite{lorenz_deterministic_1963}}{}}
The three-dimensional Lorenz system is described by the coupled differential equations
\begin{equation}
\begin{split}
    \frac{\mathrm{d}x(t)}{\mathrm{d}t} &= \sigma (y-x)\,,\\
    \frac{\mathrm{d}y(t)}{\mathrm{d}t} &= x (\rho -z) -y\,,\\
    \frac{\mathrm{d}z(t)}{\mathrm{d}t} &= xy-\beta z\,.
\end{split}
\end{equation}
Here $\sigma$, $\rho$, and $\beta$ are parameters, and $(x(t), y(t), z(t))$ describes the trajectory of the system. For $\sigma = 10$, $\rho = 28$, and $\beta = 8/3$, the system exhibits chaotic behavior. We use these values and the initial conditions $x(t=0)=y(t=0)=z(t=0)=1$. We obtain the solution from \texttt{ReservoirPy} using a Runge-Kutta method with a step size of 0.03. We drop the first 500 data points to avoid transient initialization effects.

%% file: content/appendix_05_convergence.tex
\section{Convergence criteria in training}
\label{app:convrgence}
Convergence is judged by the following criteria, which are also used in~\cite{bowles_better_2024}. At each epoch, we look at the last 400 loss values of the validation set. We compute the mean of the first 200 values $\mu_1$, the mean of the second 200 values $\mu_2$, and the standard deviation of the second 200 values $\sigma_2$. Training stops when the condition
\begin{equation}
    |\mu_1 - \mu_2| < \frac{\sigma_2}{2\sqrt{200}}
\end{equation}
is satisfied. During training, the loss decreases, so the difference between $\mu_1$ and $\mu_2$ should be greater than half the standard deviation of $\mu_2$. Once the model converges, the difference between $\mu_1$ and $\mu_2$ should be less than the standard deviation of $\mu_2$ over 200 epochs. Observing the loss over the epochs for all models, we found this criterion to be suitable to determine the convergence of the models.

%% file: content/appendix_06_scaling_seq_length.tex
\section{Scaling of the prediction error with the sequence length}
\label{app:scaling_sequ_length}
\begin{figure*}
    \centering
    \includegraphics[width=0.8\linewidth]{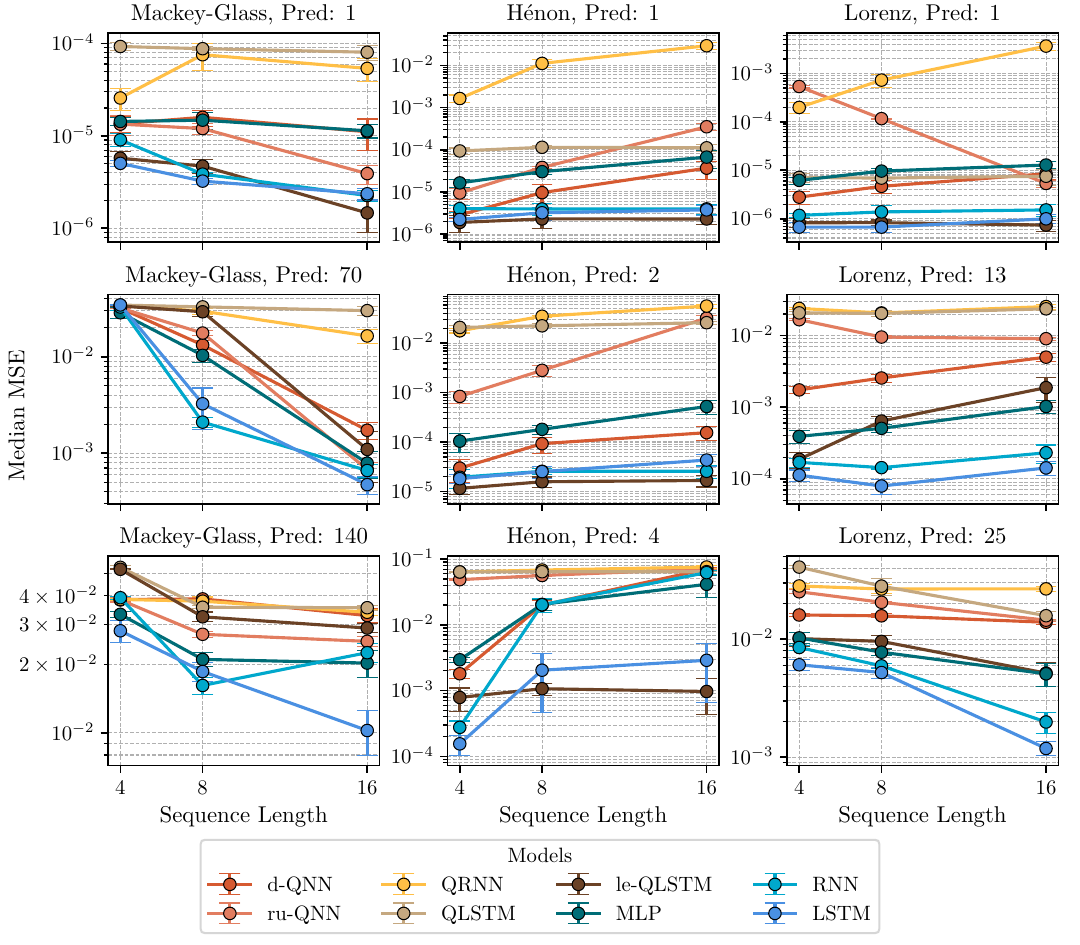}
    \caption{Median MSE on the test data set over the sequence length for different prediction tasks. Error bars represent the MAD over ten random initializations. The results reflect the best performance of all model architectures and hyperparameters tested.}
    \label{fig:scaling_seq_length}
\end{figure*}

Figure~\ref{fig:scaling_seq_length} displays the median MSE on the test data set as a function of sequence length. As discussed in the main text, we observe that for the Hénon data, the prediction error increases with longer sequence lengths. In contrast, for the Mackey-Glass and Lorenz data, the prediction error decreases. This indicates that the optimal sequence length is heavily dependent on characteristics in the data, such as the mean period and Lyapunov time. Crucially, we find that quantum and classical models exhibit qualitative similar scaling behaviors. This consistency indicates that the qualitative results of the benchmark are not artifacts of the specific sequence lengths chosen, and likely hold for larger or smaller sequence lengths.

%% file: content/appendix_07_number_parameters.tex
\section{Plots prediction accuracy over number of parameters}
\label{app:mse_over_parameters}
\begin{figure*}
    \centering
    \includegraphics[width=0.8\linewidth]{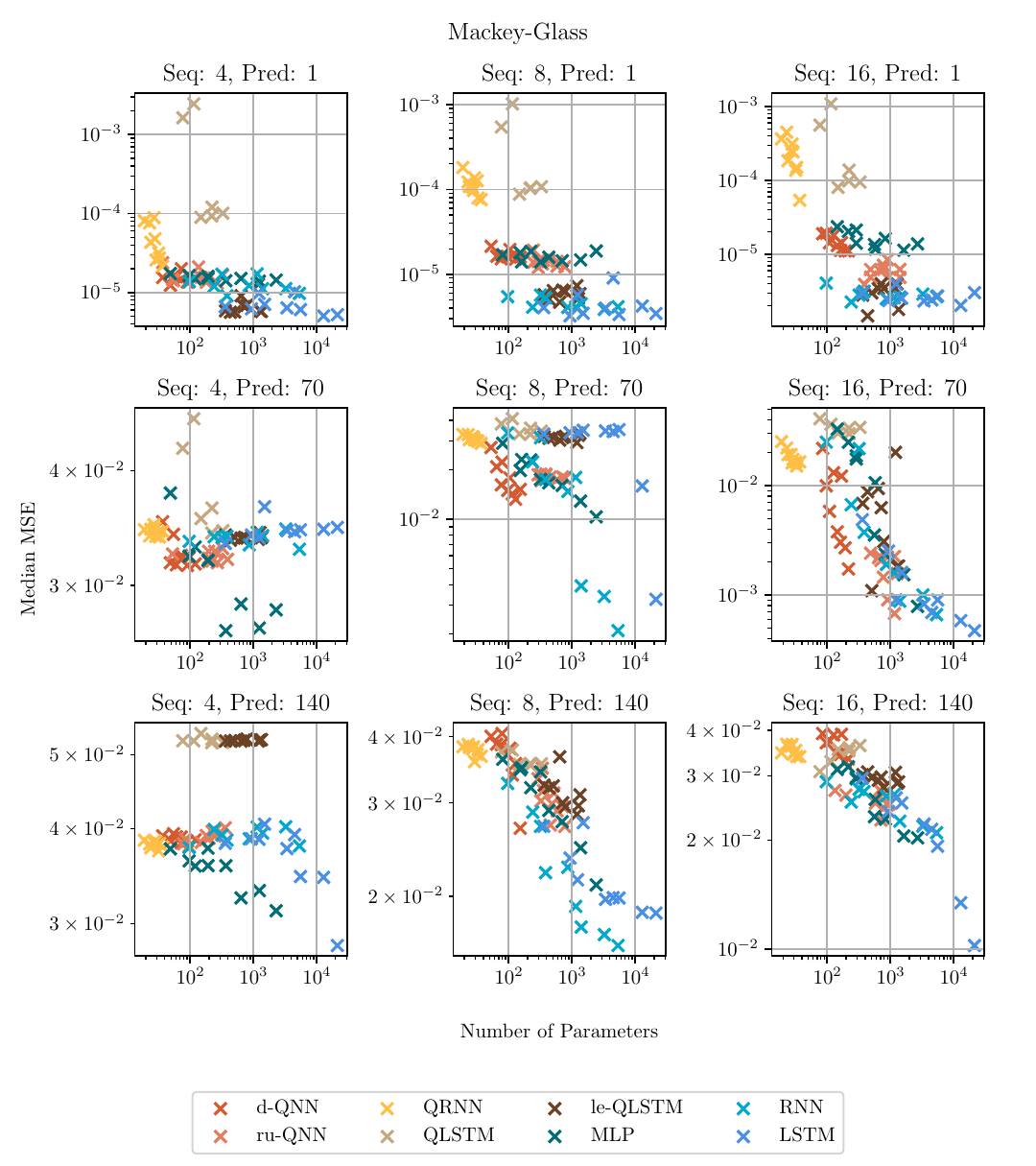}
    \caption{Median MSE on the test data set over the number of parameters for different hyperparameter configurations of the different models. Different subplots show different combinations of the parameters sequence length and prediction steps. The results shown correspond to training on the Mackey-Glass data set.}
    \label{fig:parameters_mackey}
\end{figure*}

\begin{figure*}
    \centering
    \includegraphics[width=0.8\linewidth]{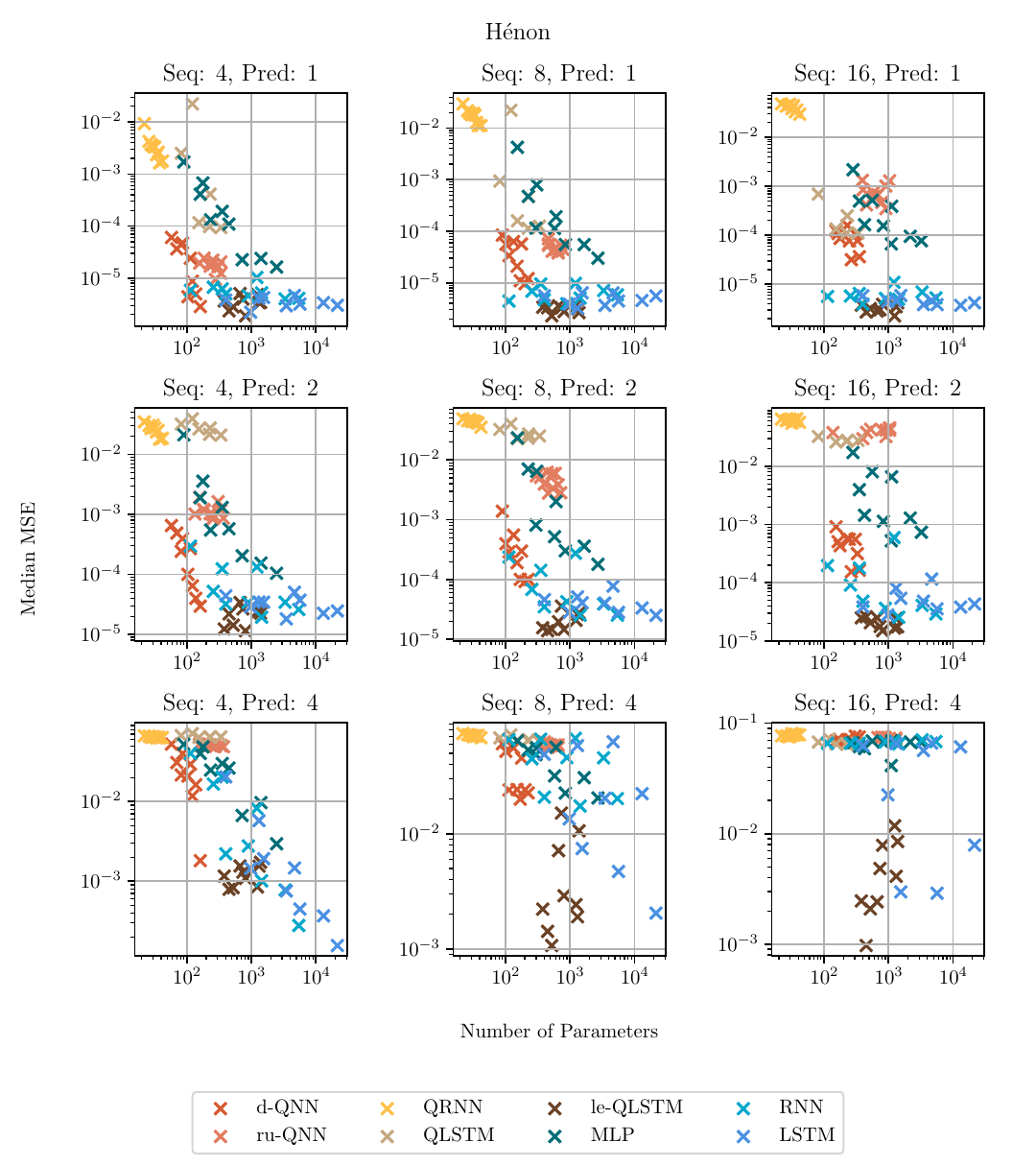}
    \caption{Median MSE on the test data set over the number of parameters for different hyperparameter configurations of the different models. Different subplots show different combinations of the parameters sequence length and prediction steps. The results shown correspond to training on the Hénon data set.}
    \label{fig:parameters_henon}
\end{figure*}

\begin{figure*}
    \centering
    \includegraphics[width=0.8\linewidth]{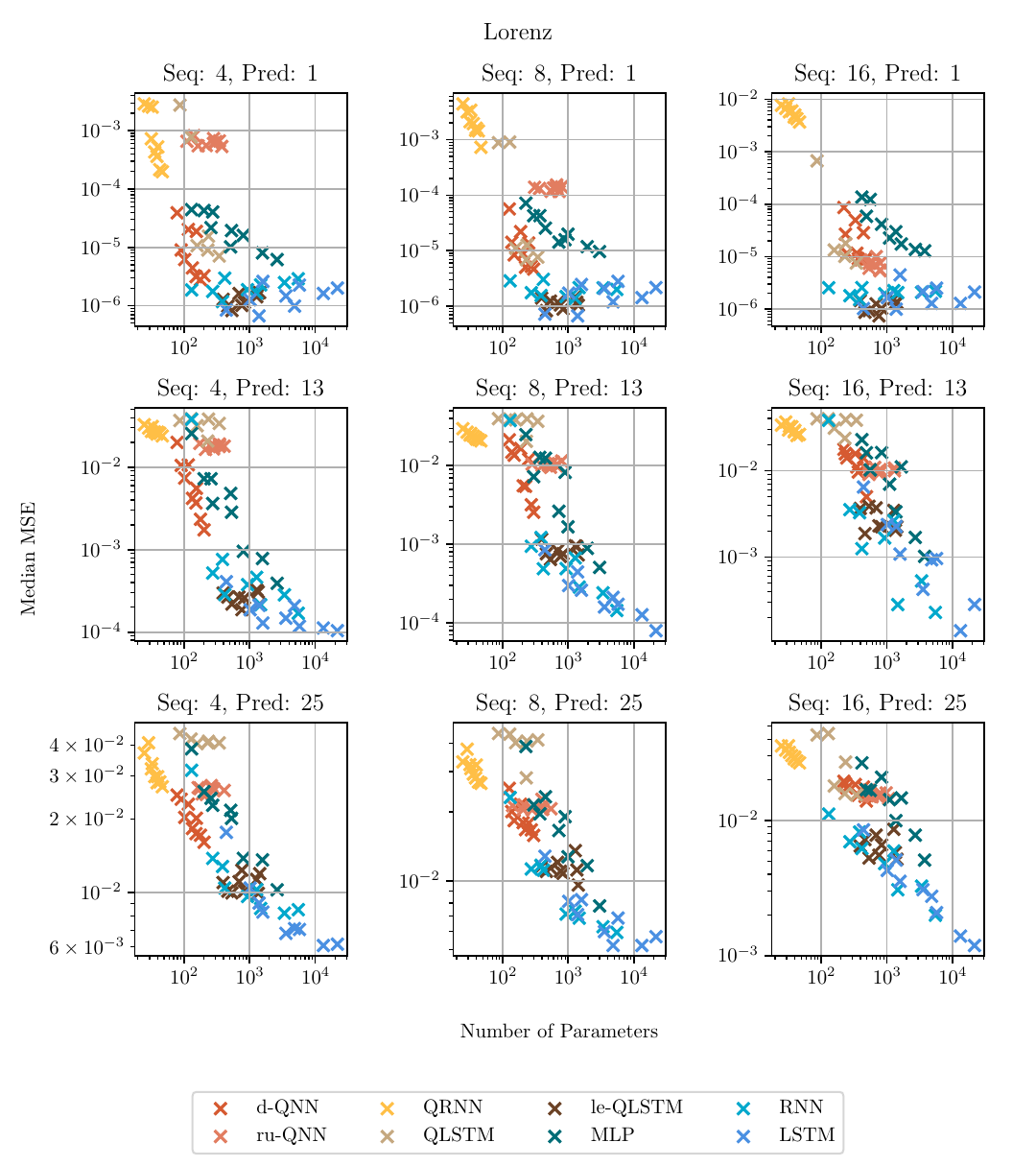}
    \caption{Median MSE on the test data set over the number of parameters for different hyperparameter configurations of the different models. Different subplots show different combinations of the parameters sequence length and prediction steps. The results shown correspond to training on the Lorenz data set.}
    \label{fig:parameters_lorenz}
\end{figure*}

In Figures~\ref{fig:parameters_mackey}-\ref{fig:parameters_lorenz} we show all plots of the median MSE on the test data set over the number of trainable parameters in the models. These figures complement Figure~\ref{fig:models_over_parameter}, which shows only two plots for the Lorenz data set, and show that the discussion in Chapter~\ref{chap:results} can be generalized to other prediction tasks. 

\label{app:parameters}

